\documentclass[aip,cha,reprint,onecolumn,showkeys,floatfix,byrevtex]{revtex4-1}

\usepackage{graphicx}
\usepackage{amsmath}
\usepackage{amssymb}
\usepackage{epsf}
\usepackage{color}

\begin{document}

\title{Interplay of synergy and redundancy in diamond motif}

\author{Ayan Biswas}
\email{ayanbiswas@jcbose.ac.in}
\affiliation{Department of Chemistry, Bose Institute, 93/1 A P C Road, Kolkata 700009, India}

\author{Suman K Banik}
\email{skbanik@jcbose.ac.in}
\affiliation{Department of Chemistry, Bose Institute, 93/1 A P C Road, Kolkata 700009, India}

\begin{abstract}
The formalism of partial information decomposition provides a number of independent components which altogether constitute the total information provided by the source variable(s) about the target variable(s). These non-overlapping terms are recognized as unique information, synergistic information and, redundant information. The metric of net synergy conceived as the difference between synergistic and redundant information, is capable of detecting effective synergy, effective redundancy and, information independence among stochastic variables. The net synergy can be quantified using appropriate combinations of different Shannon mutual information terms. The utilization of the net synergy in network motifs with the nodes representing different biochemical species, involved in information sharing, uncovers rich store for exciting results. In the current study, we use this formalism to obtain a comprehensive understanding of the relative information processing mechanism in a diamond motif and two of its sub-motifs namely bifurcation and integration motif embedded within the diamond motif. The emerging patterns of effective synergy and effective redundancy and their contribution towards ensuring high fidelity information transmission are duly compared in the sub-motifs. Investigation on the metric of net synergy in independent bifurcation and integration motifs are also executed. In all of these computations, the crucial roles played by various systemic time scales, activation coefficients and signal integration mechanisms at the output of the network topologies are especially emphasized. Following this plan of action, we become confident that the origin of effective synergy and effective redundancy can be architecturally justified by decomposing a diamond motif into bifurcation and integration motif. According to our conjecture, the presence of common source of fluctuations creates effective redundancy. Our calculations reveal effective redundancy empowers signal fidelity. Moreover, to achieve this, input signaling species avoids strong interaction with downstream intermediates. This strategy is capable of making the diamond motif noise-tolerant. Apart from the topological features, our study also puts forward the active contribution of additive and multiplicative signal integration mechanisms to nurture effective redundancy and effective synergy.   
\end{abstract}



\maketitle


\textbf{Over the years, the theory of information, as conceptualized 
by Claude Shannon, has gained much importance in 
analyzing biological phenomena. As a recent 
development, partial information decomposition (PID) has 
been proposed as an extended framework to tackle multivariate 
information processing, storage and transmission in complex animate systems. 
As an outcome of PID, one can analyze the total information content of a system 
by dividing it into a number of non-overlapping information terms 
of differing and distinct flavors. A multivariate system, therefore, 
can have unique information, synergistic information and, 
redundant information. Application of PID in biological systems, 
till now, is not much and such analysis needs to be undertaken 
to formalize the nascent framework in perspective. We chose a 
ubiquitous network motif, namely, a diamond motif to understand 
the origin of redundant and synergistic information in a signaling 
channel. Identification and information-theoretic analysis of 
sub-motifs of the diamond motif, namely, bifurcation motif and 
integration motif, play a crucial role to satisfactorily answer the 
question raised by us. The metric of our interest, the net synergy, is 
computed as the difference in synergistic and redundant 
information. Here, it is analyzed with respect to variations in a number of 
significant systemic parameters and reveals essential 
characteristics of the interlinked information components. We 
found that effective redundant information is produced when more than 
one downstream node share a common upstream node. This 
upstream node acts as a common source of fluctuations for the 
downstream nodes, thereby causing information sharing. Our 
computations also reveal that increase in effective redundant information 
concurrently increases signal-to-noise ratio or the signal fidelity 
of the information channel. We find that separation of time scales 
of input and output nodes or components modulates transmission 
of effective redundant information in the network. In addition to this, the 
strength with which the upstream component activates its 
downstream components has to be in the low to moderate 
spectrum to achieve a significant amount of redundant information. 
We have cultivated the effect of additive and multiplicative signal 
integration mechanisms, which closely approximate the biological 
realities in well-known gene regulatory networks. It has been found 
to alter the level of the net synergy throughout biologically plausible 
parametric settings. As a final step, our study proposes efficient parameter 
regimes where one of the sub-motifs overpowers its counterpart 
in contributing the desirable net synergy, in order to be comparatively more 
noise-tolerant.}


\section{Introduction}

To optimise different biophysical processes, in continuous and 
dynamic interactions with their surroundings, animate systems 
have to perform numerous and necessary computations. The 
resulting decisions made by the organisms categorically influence 
the fitness of the species in the competition for evolutionary 
selection mechanisms \cite{Thattai2001,Fraser2004,Bergstrom2004,Bialek2005,Ghosh2005,Taylor2007,Bialek2008,Tkacik2008c,Bialek2012,Bowsher2014,Levchenko2014,Mahon2014,Selimkhanov2014,Hathcock2016}. 
To identify the governing physical principles that the species 
are constrained to obey in order to achieve adaptability in a 
fluctuating environment, sophisticated measures arising from 
information theory \cite{Shannon1948,Cover1991,MacKay2002,Bialek2012} 
have been proven to be logistically handy as well as predictive \cite{Borst1999,Mitra2001,Ziv2007,Tkacik2008c,Lestas2010,Tostevin2010,Cheong2011,Tkacik2011,Rhee2012,Bowsher2013,Bowsher2014,Mahon2014,Tsimring2014,Hansen2015,Levchenko2014,Bergstrom2004}. These optimizing physical principles often dictate 
the organism to opt for certain distinctive 
architectural complexity \cite{Thattai2001,Milo2002,Mancini2013}.    
The investigation regarding the connection between the topological 
features of recurrent biological motifs and efficient information 
processing does not require detailed knowledge of the architectural 
details of the model system. This is because information 
theory deals with the biological motif as a signal communication 
system consisting message transmitting 
source, receiver at the output point and the intermediate signal 
propagation pathway. This pathway may be regarded as a black box 
where the noise comes in to 
corrupt the purity of the transmitted message 
\cite{Shannon1948,Weaver1949}.

Diamond motif (DM) is one of the recurring biological patterns 
\cite{Alon2006,Alon2007,Milo2002} and is interesting on a number of 
counts. It is one of the two prominent four node motifs found in 
signal transduction networks, the other being the bi-fan. Surprisingly, 
DM (initially known as Bi-parallel \cite{Milo2002}) is found 
in diverse networks, e.g. in neuronal networks of 
\textit{C. elegans}, ecological food webs and in forward logic 
chips embedded in electronic circuits 
\cite{Milo2002,Alon2006,Alon2007}. DM is generalized 
as multiple layered perceptrons found in the neural 
network of \textit{C. elegans} \cite{Alon2006}. One can 
decompose the DM into combinations of various sub-motifs 
and discuss the arising advantage points. We know one such 
prominent attempt using information-theoretic analysis where 
the authors dissected the DM into two 
two-step cascade (TSC) motifs and characterized profiles of gain, 
noise and the gain-to-noise ratio \cite{Ronde2012}. In that paper,  
DM is evoked in the perspective of multimerization 
and their analysis shows the emergence of a band-pass filter type 
behavior of DM. One of their key conclusions suggested
that network performance is independent 
of its architectural features and can be manipulated by modification 
of inter-species coupling strengths. Another approach has identified 
DM as a generalization of incoherent feed-forward loop motif 
and the results showed its band-pass filter type 
response to signal with temporal periodicity \cite{Cournac2009}. 
It should be noted here that the analysis performed in the existing 
literature \cite{Cournac2009,Ronde2012} used 
biologically plausible parameters due to non-availability 
of experimental data.

\begin{figure}[!t]
\begin{center}
\includegraphics[width=1.0\columnwidth,angle=0]{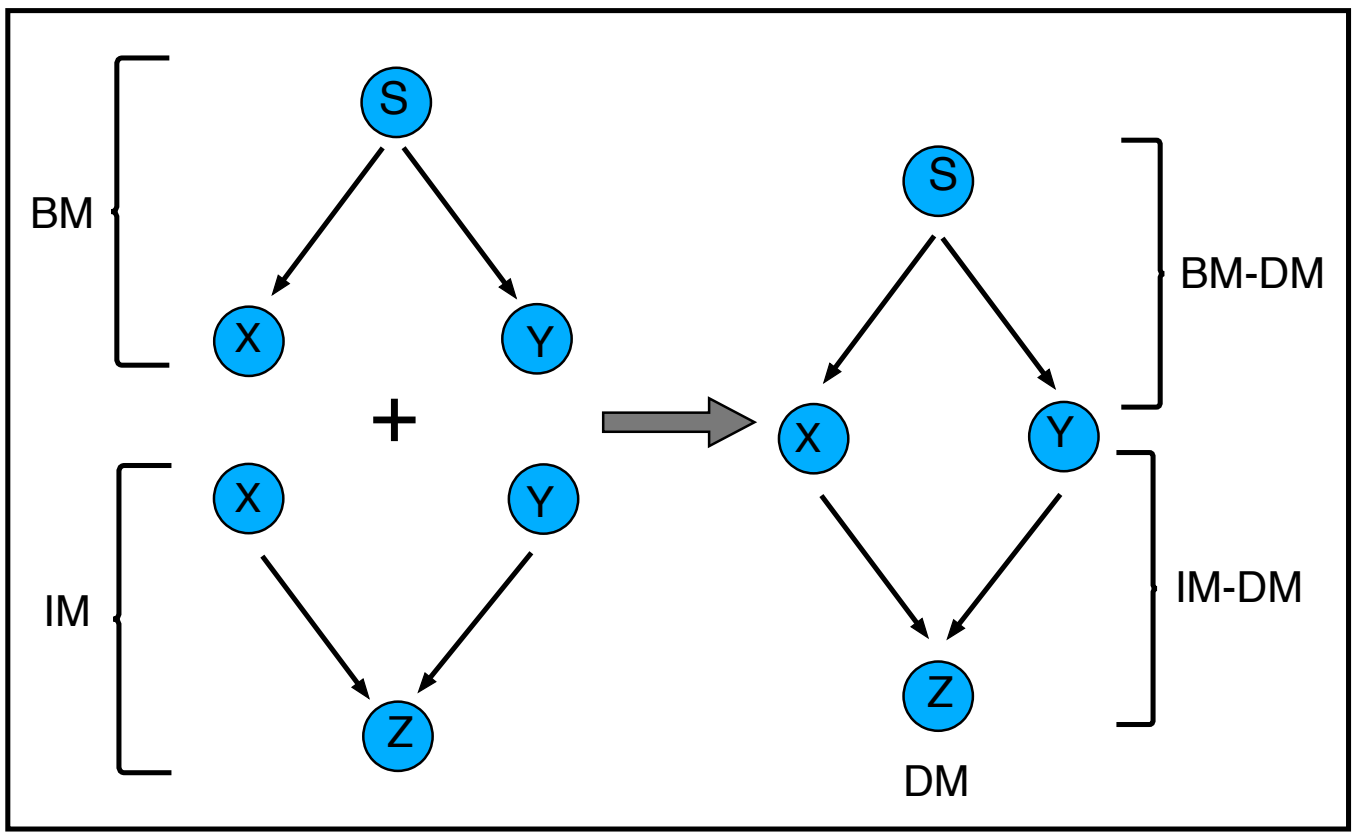}
\end{center}
\caption{Schematic diagram of bifurcation motif (BM), integration motif (IM) 
and diamond motif (DM). BM-DM and IM-DM respectively stands for the bifurcation and 
integration sub-motif embedded within the diamond motif.
}
\label{fig1}
\end{figure}

In the present communication, we have looked 
into the problem of efficient information processing in the 
DM from a fresh perspective. One can think of constructing 
DM using independent bifurcation motif (BM) and independent 
integration motif (IM). It is also possible to identify 
two different sub-motifs embedded in DM, namely bifurcation 
sub-motif (BM-DM) and integration sub-motif (IM-DM) 
(see Fig.~\ref{fig1}). 
We have adopted an information-theoretic measure which 
involves a suitable combination of three-variable and 
two-variable mutual information (MI) terms and the 
combination is known as the net synergy 
\cite{Schneidman2003,Barrett2015}. It has the 
potential to predict synergy, redundancy and, information 
independence among stochastic variables involved in 
information processing in a complex system. The net synergy is a
product of partial information decomposition (PID) \cite{Williams2010,Barrett2015}. 
Using PID, we can decompose the total information provided by a 
set of source variables about a target variable into 
independent or non-overlapping information terms namely 
unique information, synergistic information and, redundant 
information. Unique information is the information about the 
target variable provided only by a specific source variable 
whereas redundant information about the target is provided 
by all the source variables holding common shares. Synergistic 
information about the target is provided jointly by the set of source variables. 
In other words, to get hold of the synergistic information about 
a target variable, one has to know all the source variables 
simultaneously.  As mentioned by Barrett \cite{Barrett2015}, 
for a system of three random variables, one can link three 
MI terms with four information terms generated through PID (i.e. two unique 
information, synergistic information and, redundant information) 
using only three equations. Four unknown PID terms related 
through three equations make the system under-specified in 
the absence of any specific definition for any one of the unknowns. 
Hence, from this system of equations, only the difference between 
synergistic information and redundant information (i.e. the net synergy) may be obtained
\cite{Barrett2015}. The positive value of net synergy implies 
that synergy is dominant over redundancy. For negative 
values of the net synergy, redundancy overpowers synergy. The 
borderline case of zero net synergy indicates information 
independence among information source and target variables 
\cite{Schneidman2003,Barrett2015}.

There exists rich literature on multivariate 
information decomposition which is an active field of research. 
For a comparative discussion on different measures involving multivariate information 
along with their applications, the review by Timme \textit{et. al.,} 
is notable \cite {Timme2014}. In a broader perspective, Faes 
\textit{et. al.,} have been able to show that computation of 
information storage and its transfer are necessary to predict 
the dynamics of target variable \cite{Faes2015a}. Synergistic and 
redundant information transfer components have been exactly
computed for coupled Gaussian systems operating across
multiple time scales \cite{Faes2017b}. It has been also pointed out 
that the questions regarding definitions of what constitutes 
synergy and redundancy in case of nonlinear systems are at present
not satisfactorily settled\cite{Faes2017b}.
Apart from information storage and transfer, Ref.~\onlinecite{Faes2017}, 
directs our attention to assemble information modification 
formalism to better understand dynamics of complex network 
topologies. Besides, composite analysis of information-dynamic 
measures can underpin the causal effects that follow from the 
dynamics which itself is susceptible to varying experimental 
conditions \cite{Faes2015b}. In a study performed by Bertschinger 
\textit{et. al.,} measures have been proposed for decompositions 
of multivariate MI-s along with a working concept for the unique 
information \cite{Bertschinger2014}. Different quantifiers of 
synergistic information have been applied in a set of binary 
circuits for their comparative analysis \cite{Griffith2014}. For a 
new formalism regarding redundant information and its usage 
to decompose transfer entropy, one can take note of the proposal 
by Harder \textit{et. al.,} \cite{Harder2013}. Recent findings 
according to Wibral and colleagues have projected PID as a 
consistent framework such that it can compare 
different neural goal functions and formulate potential new 
candidates \cite{Wibral2017}. To supplement advances in 
theoretical understanding on multivariate information 
decomposition, experimentalists have also contributed their 
fair share in this ever-growing research domain. Gawne and 
Richmond have experimented on synergistic, redundant and, 
independent information encoding inside the inferior temporal 
cortex region in behaving rhesus monkey brain \cite{Gawne1993}. 
The role played by correlations in the encoding mechanisms 
inside the nervous system has also been central to elaborate 
and thought-provoking research \cite{Panzeri1999}. Brenner \textit{et. al.,} demonstrated 
that even the motion stimulated visual information processing device of a fly uses 
synergistic code \cite{Brenner2000}. Analytical study utilizing
information theory and focussing on the importance of correlations
related to somatosensory cortical populations in rats can be found in 
Ref.~\onlinecite{Montani2009}. This interesting paper has presented 
the fact that to actually explain their experimental data, it is indeed
important to consider correlational order greater than two.

In the present work, we have not attempted any extension 
or modification of the existing formalism of PID. Instead, we 
have a clear objective to apply PID to study information 
processing in a ubiquitous gene regulatory network (GRN) motif, i.e. 
DM. Barrett \cite{Barrett2015} has studied some example 
cases of dynamical systems having biological analogs, 
but the processes are modeled as  of multivariate 
autoregressive nature. It makes the time-evolution of the 
system discrete and the interactions linear, which are 
seldom shown by living systems. We envisaged the 
situation quite the opposite, i.e. a system with continuous 
temporal variations and nonlinear interactions.
The discreteness in time is brought by the 
author to showcase the effect of past states of the source 
variables on the future state of the target variable. However, here 
we are concerned with the net synergy aspects where all 
the variables are at the same time point, i.e. without any 
time-lag.

The role of redundancy to combat noise in the 
information channel was previously 
presented by Shannon and Weaver in their illuminating 
book \cite{Shannon1963} which motivated us 
to look for analogs in model network 
motifs.  The theoretical explanation regarding the origin of 
redundancy in a two-step cascade motif and its 
connection with information fidelity have been discussed 
thoroughly in our earlier work \cite{Biswas2016}. In 
Ref.~\onlinecite{Tkacik2009}, authors have presented a 
case of redundancy where multiple genes with successively 
higher values of activating signal strengths, are driven by 
a single input. Another interesting recent study has found 
out that genetic redundancy along with intrinsic noise and 
heterogeneity can increase information transfer whereas 
extrinsic noise and cross-talk have an inverse effect 
\cite{Rodrigo2016}. Immunofluorescence readouts from 
network experiments involving NF-$\kappa$B and ATF-2  
receiving the signal from tumor necrosis factor (TNF) through 
TNF receptor, reveal the mitigation of noisy effects. Thus information
propagation is enhanced with the help of redundancy \cite{Cheong2011}. 
Rhee and colleagues have been successful in detecting connections between 
network architecture and associated noise of biochemical 
origin \cite{Rhee2014}. Whereas these studies have 
reported the connection between information fidelity and 
redundancy (architectural or/and informational), our 
analysis has quantified this connection in model network 
motif with a strong emphasis on PID. Moreover, in our 
current initiative, attempts have been made to showcase 
the connection between network topology and the 
synergy-redundancy duo. Also, to be specific, this line of 
analysis qualifies as being the central theme of the 
present report.

To this end, we have chosen signal-to-noise ratio (SNR) 
as the measure of fidelity \cite{Bowsher2013,Biswas2016} 
in the motifs. Gain-to-noise ratio (GNR) plays another strong 
candidate which can successfully quantitate the performance 
of the network motif, and it does so independently of the signal 
characteristics \cite{Tostevin2010}. One can also link GNR to 
the Fisher information about the signaling species provided 
by the output species or response \cite{Cover1991,Tostevin2010}. 
The analysis performed here has been done at steady states 
of involved species as it is suggested that living systems 
perform optimally in their steady states and keeping 
concentrations or copy numbers fixed helps to compare 
different parametric scenarios on an equal footing \cite{Alon2006}. 
Additionally since in our case, the signaling species driving 
the DM follows Poisson process, SNR is found to be 
proportional to GNR. Here, the steady state ensemble 
averaged population of the signaling species serves as the 
constant factor of proportionality. 


We have used the relaxation time scale as a parameter 
because the separation of time scale is crucial to dictate the  
information flow in a biochemical network behaving 
like a  noise filter 
\cite{Bruggeman2009,Tostevin2010,Ronde2012,Hinczewski2014,Maity2015,Hathcock2016}. 
To this end, it is important to mention that the biological 
system's reaction to the input signal is indeed amplitude 
and frequency dependent. Since, DM and the associated 
motifs and sub-motifs are the living analogs of an electronic 
signal processing device, concepts of encoding and 
decoding can be equivalently mapped to these biological 
motifs \cite{Schneidman2003, Hansen2015}. 
Here, we show that the features of the net synergy 
landscape of the motifs and sub-motifs are intricately 
connected to the principles of communication theory.

The response of a transcription control network 
depends on the mechanism adopted by the 
organism to integrate multiple incoming signals. 
For example, in the well-characterized 
\textit{lac} system of \textit{E. coli}, this integration 
machinery is intermediate between boolean OR logic 
and AND logic. By implementing a few point mutations in 
the system, pure logic such as OR, AND or, single-input 
switches have been achieved. It implies the plasticity of 
the \textit{lac} operon's integration function 
\cite{Setty2003,Mayo2006}. These fundamental discoveries 
motivated us to investigate how does the signal integration 
mechanism affect the characteristics of the net synergy 
profiles and vary the fidelity of signal processing. 
Though in principle, a significant number of different logic gates can be implemented in 
synthetic set-ups, evolvability of these modular structures 
inside living cells is not always guaranteed 
\cite{Silva-Rocha2008}. In this connection, we find PID 
to provide some preliminary ideas. For the sake of 
simplicity, we restrict the input integration mechanism 
to be either additive or multiplicative, keeping the 
expression levels of all  the biochemical species 
constant throughout different cases for better comparison.

Some other technical points that the present report 
takes benefit of are as follows. We have used Gaussian 
random variables to represent the biochemical species 
so that we can treat the MI of various channels as their 
respective channel capacities \cite{Tostevin2010} and 
these can be easily computed using corresponding 
variance and covariance. We have made our analysis 
tractable further by assuming Gaussian noise processes 
\cite{Ziv2007,Tostevin2010,Maity2015,Bruggeman2009,Ronde2010,Bialek2005,Tanase2006,Warren2006,Maity2014,Grima2015,Biswas2016,Tostevin2010}. 
We use noise with zero cross-correlation, because 
biological phenomenology dictates the validity of this 
approach in a birth-death type of dynamics \cite{Ronde2012}, 
as dealt here. Another vital point to note here is the 
consideration of low copy number 
\cite{Elowitz2002,Paulsson2004,Bialek2005,Bialek2008,Tkacik2009} 
of the biochemical species involved in the stochastic 
reactions.

We performed the theoretical analysis 
using linear noise approximation (LNA) 
\cite{Elf2003,Tanase2006,Kampen2007,Tostevin2009,Ronde2010,Grima2015} 
to handle nonlinearity which enters into the system 
through the Hill type regulatory functions 
\cite{Bintu2005,Ziv2007,Tkacik2008b,Tkacik2009,Walczak2010} 
used in our mathematical model. We do take note of the 
fact that there are other techniques, e.g. the small-noise 
approximation that can be applied to tackle the fluctuations 
in the system \cite{Tkacik2008a,Tkacik2009}. Though low 
copy numbers of the reacting molecules contribute a 
significant amount of noise into the reaction volume, a 
good match between our analytical results and stochastic 
simulations based on Gillespie's method 
\cite{Gillespie1976,Gillespie1977} establishes the validity 
of LNA. In this connection, our observations receive 
strong support from previous findings of similar type 
considering copy number as low as $\sim 10$. 
\cite{Ziv2007,Bruggeman2009,Maity2015,Biswas2016} 
From current literature \cite{Grima2015}, LNA is known 
to be effective beyond high copy number 
reactions. It gives out exact results up to the second 
moments of the system components involved in 
second-order reactions. There may be a number of 
species that are involved in these specific type of 
reactions, but it is noted that at least one of those 
reacting species in every such reaction fluctuates 
not only in a Poissonian way but also in an 
uncorrelated fashion with the rest of the species 
\cite{Grima2015}.


\section{The Model}

The set of  Langevin equations governing the dynamics of a DM are,
\begin{eqnarray}
\label{eq1}
\frac{ds}{dt} & = & f_{s}(s)-\mu_s s+\xi_s(t), \\
\label{eq2}
\frac{dx}{dt} & = & f_{x}(s,x)-\mu_x x+\xi_x(t), \\
\label{eq3}
\frac{dy}{dt} & = & f_{y}(s,y)-\mu_y y+\xi_y(t), \\
\label{eq4}
\frac{dz}{dt} & = & f_{z}(s,x,y,z)-\mu_z z+\xi_z(t).
\end{eqnarray}

\noindent Here, to represent the copy numbers of species S, X, Y 
and, Z present in the unit amount of cellular volume, we use the 
symbols $s, x, y$ and, $z$, respectively. To be precise, if one 
considers these biochemical species to be transcription factors, 
the corresponding volume has to be an effective volume since 
these transcription factors after being produced in the cytoplasm 
are carried inside the nucleus where they are sensed. Hence 
merely considering either the cellular or the nuclear volume would 
be inaccurate \cite{Walczak2010}. The set of Langevin equations 
written above is suggestive of birth-death type of mechanisms 
governing the population levels. We have modelled degradation 
to be proportional to the respective population size with 
$\mu_{i}$-s ($i = s,x,y,z$) setting the time scale of degradation. 
The inter-species interactions are manifested through the synthesis 
of X, Y and, Z. These terms are taken to be nonlinear, in general, 
in agreement with real biological scenario \cite{Bintu2005,Ziv2007,Tkacik2008a,Tkacik2008b,Ronde2012}. 
In the present analysis we use $\langle \xi_{i}(t) \rangle$ = 0 and 
$\langle \xi_{i}(t)\xi_{j}(t^{'}) \rangle$ = $\langle |\xi_{i}|^{2}\rangle \delta_{ij} \delta(t-t^{'})$, 
which makes the noise processes independent and Gaussian 
distributed. At steady state the noise strength becomes 
$\langle |\xi_{i}|^{2}\rangle = \langle f_{i} \rangle + \mu_{i}\langle i \rangle = 2\mu_{i}\langle i \rangle$ where $i =s, x, y$ and, $z$ 
\cite{Elf2003,Swain2004,Paulsson2004,Tanase2006,Warren2006,Kampen2007,Mehta2008,Ronde2010}. The first equality demonstrates that both synthesis and degradation processes are sources for noise in the system 
and their individual contributions add up to produce the ultimate 
steady state noise strength. The second equality indicates that both the noise 
sources contribute in equal proportions at steady state. 
The usage of $\langle \cdots \rangle$ denotes steady state ensemble 
average over many independent realizations. To calculate the second 
moments of $s, x, y$ and, $z$ through LNA, we apply perturbation 
of linear order $\delta u(t) = u(t) - \langle u \rangle$ with 
$\langle u \rangle$ being the average population of $u$ at steady 
state and recast Eqs.~(1-4) in the following form
\begin{equation}
\label{eq5}
\frac{d\mathbf{\delta W}}{dt} = \mathbf{J}_{W=\langle W \rangle}
\mathbf{\delta W}(t) + \mathbf{\Xi}(t),
\end{equation}

\noindent 
where we denote by $\mathbf{\delta W}(t)$, the fluctuation matrix 
containing the linear order perturbations and the noise matrix by 
$\mathbf{ \Xi}(t)$
\begin{eqnarray*}
\mathbf{\delta W}(t) = \left( 
\begin{array}{ccc} \delta s(t) \\ \delta x(t) \\ \delta y(t) \\ \delta z(t) \\
\end{array} \right),
\mathbf{ \Xi}(t) = \left( 
\begin{array}{ccc}  \xi_{s}(t) \\  \xi_{x}(t) \\  \xi_{y}(t) \\  \xi_{z}(t) \\
\end{array} \right),
\end{eqnarray*}

\noindent
$\mathbf{J}$ represents the Jacobian matrix at steady state. 
The Lyapunov equation at steady state 
\cite{Keizer1987,Elf2003,Paulsson2004,Paulsson2005,Kampen2007}
\begin{equation}
\label{eq6}
\mathbf{J \Sigma} + \mathbf{\Sigma J}^{T} + \mathbf{D} = \mathbf{0},
\end{equation}

\noindent establishes connections between the steady-state 
fluctuations of the biochemical species and noise-driven 
dissipation in the system. The fluctuation part is encapsulated 
in $\mathbf{\Sigma}$ which is the covariance matrix, and 
$\mathbf{D}$ contains the dissipation part since its entries are 
various noise strengths, i.e. 
$\mathbf{D}$ = $\langle \mathbf{\Xi}  \mathbf{\Xi}^{T} \rangle$ 
where $T$ denotes matrix transposition operation.

In the generalized analytic expressions of the second moments 
(see Appendix) obtained by solving the Lyapunov equation at 
steady state, $s,x,y$ and, $z$ are approximated as 
$\langle s \rangle, \langle x \rangle, \langle y \rangle$ and, 
$\langle z \rangle$, respectively \cite{Ronde2010,Maity2015,Biswas2016}. 
These second moments serve as the ingredients for calculating 
the two-variable and three-variable MI terms. For computing the 
net synergy (in the unit of `bits') among S, X and, Y which 
constitute BM and BM-DM, and among X, Y and, Z which 
constitute IM and IM-DM, we use the following two expressions 
\cite{Schneidman2003,Williams2010,Barrett2015,Hansen2015,Biswas2016}
\begin{eqnarray}
\label{eq7}
\Delta I(s;x,y) & = & I(s;x,y)-I(s;x)-I(s;y), \\
\label{eq8}
\Delta I(z;x,y) & = & I(z;x,y)-I(z;x)-I(z;y).
\end{eqnarray}

\noindent Here, it is to be kept in mind that information-theoretic 
characterization of the random variables may not be at par with 
the causal relations among them \cite{Barrett2015}. Though 
S regulates X and Y, which in turn regulate Z, it is feasible to 
consider X and Y as the information source and S as the information target while 
computing $\Delta I(s;x,y)$. In contrast, we regard X and Y as the 
information source and Z as the information target to calculate $\Delta I(z;x,y)$. The choice 
of grouping variables into these categories should be dictated 
by the rationale of the question being asked for specific 
network architecture. In the framework of Shannon, the network 
motifs are logically thought of as signal processing channels, 
and by the same token, the information propagation phenomenon 
can be described as a joint encoding-decoding process 
\cite{Shannon1948,Schneidman2003}. $\Delta I(s;x,y)$ gets the 
advantage of the decoding scheme taking the intermediates 
X and Y as the source of information for S. On the other side, these 
same intermediates can take part in encoding information and 
transfer to Z and hence, the relevant net synergy turns out to 
be $\Delta I(z;x,y)$. Specifically in our system, one of the key 
questions that we want to investigate is how does a common 
source of fluctuations affect the net synergy landscape. Since 
the intermediate information hubs X and Y are common nodes 
in the sub-motifs, one can infer both S (decoding) and Z (encoding) 
starting with X and Y. We emphasise that the PID induced 
source-target specifications do not levy any constraint on the 
chemical kinetics of these species.

According to Ref.~\onlinecite{Barrett2015}, synergistic information 
($I_S$) and redundant information ($I_R$) constitutes the net 
synergy $\Delta I$ as follows
\begin{eqnarray}
\label{eq9}
\Delta I & = & I_S - I_R.
\end{eqnarray}

\noindent It is evident from the above equation that positive 
net synergy ($\Delta I >0$) reveals a greater amount of 
synergy in comparison with redundancy while negative net 
synergy ($\Delta I <0$) reverses the situation. One should 
take note of the fact that $\Delta I >0$ arises when 
$I_S > I_R \ge 0$, i.e. $I_R$ may or may not exist but $I_S$ 
is sure to exist and is greater in proportion than $I_R$. On 
the other hand, $\Delta I <0$ implies $I_R > I_S \ge 0$, i.e. 
$I_S$ may or may not be there but $I_R$ is definitely present 
and is more significant than $I_S$. For the sake of clarity, we paraphrase 
$\Delta I >0$ by effective synergy $I_{ES}$ and $\Delta I <0$ 
by effective redundancy $I_{ER}$. Synergy and redundancy 
can also balance each other, and this gets reflected in zero 
amount of the net synergy ($\Delta I =0$). This situation 
demands $I_S=I_R \ge 0$, i.e. either both $I_S$ and $I_R$ 
exist in equal proportion or both are absent. One should be 
mindful about these conceptual points while analyzing the 
net synergy profiles generated with respect to variations of 
different system parameters and should not be led to infer 
that any attempt to quantify absolute synergy and absolute 
redundancy has been made. The existence of pure synergy and 
pure redundancy have been speculated entirely based upon 
the qualitative nature of synergistic and redundant information 
due to the framework of PID.

Also, one should not conclude based on the source-target 
classifications and the polarity of the net synergy that these 
metrics provide directionality of information flow. MI is a 
sophisticated and generalized measure of correlation, and 
it is strictly semi-positive. The net synergy also captures 
correlation more finely with respect to information 
independence of the associated random variables 
\cite{Schneidman2003}. Besides, correlation does not infer 
causation. For that matter, since we are considering a 
directed network, the direction of information flow is entirely 
determined by the specified interactions. Now, for Gaussian 
random variables, Eqs.~(\ref{eq7}-\ref{eq8}) become 
\cite{Barrett2015}
\begin{eqnarray}
\Delta I(s;x,y) & = & \frac{1}{2} \left (
\log_{2} \left  [ \frac{\det \Sigma(s)}{\det \Sigma(s|x,y)} \right ] - 
\log_{2} \left [ \frac{\det \Sigma(s)}{\det \Sigma(s|x)} \right ] \right.
\nonumber \\
\label{eq10}
&& \left. - \log_{2} \left [ \frac{\det \Sigma(s)}{\det \Sigma(s|y)} \right ]
\right ), \\
\Delta I(z;x,y) & = & \frac{1}{2} \left (
\log_{2} \left  [ \frac{\det \Sigma(z)}{\det \Sigma(z|x,y)} \right ] - 
\log_{2} \left [ \frac{\det \Sigma(z)}{\det \Sigma(z|x)} \right ] \right.
\nonumber \\
\label{eq11}
&& \left. - \log_{2} \left [ \frac{\det \Sigma(z)}{\det \Sigma(z|y)} \right ]
\right ).
\end{eqnarray}

\noindent 
The consecutive terms in the right-hand side of Eq.~(\ref{eq10}) 
denote $I(s;x,y)$, $I(s;x)$ and, $I(s;y)$, respectively. Similarly, 
the consecutive terms in the right-hand side of Eq.~(\ref{eq11}) 
denote $I(z;x,y)$, $I(z;x)$ and, $I(z;y)$, respectively. The terms 
appearing in the denominators of the MI-s present in 
Eqs.~(\ref{eq10},\ref{eq11}) are the corresponding conditional 
variances and can be computed as follows \cite{Barrett2015}
\begin{eqnarray}
\label{eq12}
\Sigma(s|x) & =: & \Sigma(s)-\Sigma(s,x)[\Sigma(x)]^{-1}\Sigma(x,s), \\
\label{eq13}
\Sigma(s|y) & =: & \Sigma(s)-\Sigma(s,y)[\Sigma(y)]^{-1}\Sigma(y,s), \\
\Sigma(s|x,y) & =: & \Sigma(s) - 
\left( \begin{array}{ccc} 
\Sigma(s,x) & \Sigma(s,y)\\
\end{array} \right) \nonumber \\
\label{eq14}
&& \times 
\left( 
\begin{array}{ccc} \Sigma(x) & \Sigma(x,y)\\
\Sigma(y,x) & \Sigma(y) \\
\end{array} 
\right)^{-1} 
\left( 
\begin{array}{ccc} \Sigma(x,s)\\
\Sigma(y,s)\\
\end{array} 
\right).
\end{eqnarray}

We note that the covariances are symmetric, e.g. 
$\Sigma(s,x)$ = $\Sigma(x,s)$ etc. Similarly, one can 
compute necessary expressions involving $x$, $y$ and, $z$. 
For explicit generalized forms of the variance and covariance, 
we refer to the Appendix.


\begin{figure}[!t]
\begin{center}
\includegraphics[width=1.0\columnwidth,angle=0]{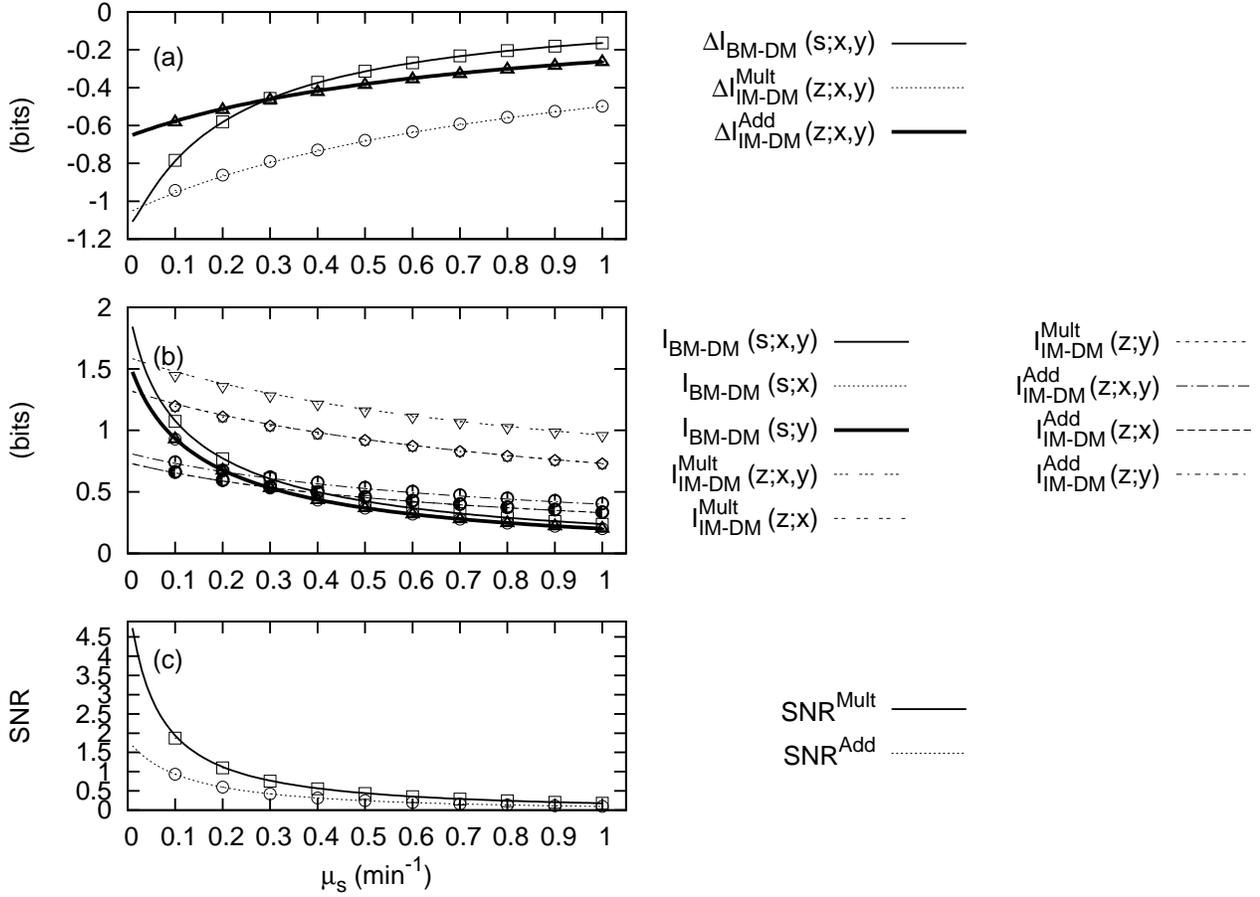}
\end{center}
\caption{(a) The net synergy for BM-DM ($\Delta I_{BM-DM} (s;x,y)$), IM-DM ($\Delta I_{IM-DM}^{Mult} (z;x,y)$ and $\Delta I_{IM-DM}^{Add} (z;x,y)$), (b) Different three variable and two variable MI for BM-DM and IM-DM, (c) SNR$^{Mult}$ and SNR$^{Add}$ as functions of $\mu_{s}$. The lines are due to analytical results and the symbols represent numerical data generated using Gillespie's algorithm \cite{Gillespie1976,Gillespie1977} with an ensemble averaging over $10^6$ independent time series. The constraint relations for species populations are $\langle s \rangle=10$, $\langle x \rangle=100$, $\langle y \rangle=100$ and, $\langle z \rangle=100$ all in molecules/$V$ with $V$ being the unit effective cellular volume. The rate parameters relevant for BM-DM and IM-DM are $k_s$ = $\mu_s \langle s \rangle$, 
$k_x$ = $\mu_x \langle x \rangle ((K_1^n+\langle s \rangle^n)/\langle s \rangle^n)$, 
$k_y$ = $\mu_y \langle y \rangle ((K_2^n+\langle s \rangle^n)/\langle s \rangle^n)$,
$k_z$ = $\mu_z \langle z \rangle (\frac{\langle x \rangle^n}{K_3^n+\langle x \rangle^n}+\frac{\langle y \rangle^n}{K_4^n+\langle y \rangle^n})^{-1}$ and, 
$k_z'$ = $\mu_z \langle z \rangle 
((K_3^n+\langle x \rangle^n)/\langle x \rangle^n)
((K_4^n+\langle y \rangle^n)/\langle y \rangle^n)$. $K_1=K_2=90$, $K_3=K_4=100$ all in molecules/$V$ with $V$ being the unit effective cellular volume, $\mu_{x}=\mu_{y}=0.5$ min$^{-1}$, $\mu_{z}=5$ min$^{-1}$. In both the cases, we have used $n=1$.  
}
\label{fig2}
\end{figure}


\section{Results and Discussion}

In this section, we explore the parametric dependence of information transmission 
in terms of the net synergy in the DM. To this end, we have identified two 
sub-motifs namely the BM-DM and the IM-DM, embedded within the DM 
(see Fig.~\ref{fig1}). The net synergy is explored in the independent BM 
and IM (see Fig.~\ref{fig1}) and also when both the motifs are ingrained in 
the DM as sub-motifs. This has been done with an aim to quantify the 
dependance of DM on two of its sub-motifs in the context of information 
processing and mitigation of noise. In both the analytical and numerical 
calculations, we keep the species population fixed at steady state e.g., 
$\langle s \rangle$ = 10, $\langle x \rangle$ = 100, $\langle y \rangle$ = 100 
and, $\langle z \rangle$ = 100, irrespective of the network architecture 
examined. We note here that the unit of the population we use is molecules/$V$,
where $V$ is unit effective cellular volume. The steady state population level
we use, are experimentally observed in transcription control network 
\cite{Paulsson2004,Golding2005,Tkacik2008b}. The strategy of copy number 
constancy has the advantage of comparing the optimal performances of species 
at steady state \cite{Alon2006}. The constraint of population constancy governs 
choices of the synthesis rate parameters while one can make independent 
choices for the degradation rate parameters.


\subsection{Variation in the time scale of input signal shows effective 
redundancy empowers signal fidelity}

Fig.~\ref{fig2} shows the net synergy $\Delta I_{BM-DM} (s;x,y)$ and 
$\Delta I_{IM-DM} (z;x,y)$ for BM-DM and IM-DM, respectively, along with 
corresponding two-variable and three-variable MI and the SNR as a function 
of $\mu_{s}$ for $\mu_{x} = \mu_{y} = 0.5$ min$^{-1}$ and 
$\mu_{z} = 5$ min$^{-1}$.  $\Delta I_{IM-DM} (z;x,y)$, MI-s and, SNR-s are
theoretically calculated for additive and multiplicative integration and denoted 
by $\Delta I_{IM-DM}^{Add}$, $\Delta I_{IM-DM}^{Mult}$, $ I^{Add} (z;x,y)$ 
and $ I^{Mult} (z;x,y)$ etc., SNR$^{Add}$ and, SNR$^{Mult}$, respectively.  
To check the validity of our analytical calculations, we also execute numerical 
simulation using stochastic simulation algorithm \cite{Gillespie1976,Gillespie1977}. 
Table~I provides the detailed information on the reaction channels implemented 
in stochastic simulation. 
For example, in the BM 
$k_s$ = $\mu_s \langle s \rangle$, 
$k_x$ = $\mu_x \langle x \rangle ((K_1^n + \langle s \rangle^n)/\langle s \rangle^n)$ and, 
$k_y$ = $\mu_y \langle y \rangle ((K_2^n + \langle s \rangle^n)/\langle s \rangle^n)$. 
Similarly for the IM we use, 
$k_x$ = $\mu_x \langle x \rangle$, 
$k_y$ = $\mu_y \langle y \rangle$, 
$k_z$ = $\mu_z \langle z \rangle (\frac{\langle x \rangle^n}{K_3^n+\langle x \rangle^n}+\frac{\langle y \rangle^n}{K_4^n+\langle y \rangle^n})^{-1}$ and, 
$k_z'$ = $\mu_z \langle z \rangle 
((K_3^n+\langle x \rangle^n)/\langle x \rangle^n)
((K_4^n+\langle y \rangle^n)/\langle y \rangle^n)$.
Finally for the DM the synthesis rate parameters are as follows:
$k_s$ = $\mu_s \langle s \rangle$, 
$k_x$ = $\mu_x \langle x \rangle ((K_1^n+\langle s \rangle^n)/\langle s \rangle^n)$, 
$k_y$ = $\mu_y \langle y \rangle ((K_2^n+\langle s \rangle^n)/\langle s \rangle^n)$, 
$k_z$ = $\mu_z \langle z \rangle (\frac{\langle x \rangle^n}{K_3^n+\langle x \rangle^n}+\frac{\langle y \rangle^n}{K_4^n+\langle y \rangle^n})^{-1}$ and, 
$k_z'$ = $\mu_z \langle z \rangle 
((K_3^n+\langle x \rangle^n)/\langle x \rangle^n)
((K_4^n+\langle y \rangle^n)/\langle y \rangle^n)$.
We note that while generating the analytical and numerical results in 
Fig.~\ref{fig2}, we have used $n=1$ as $n > 1$ brings in higher order
nonlinearity that may cause breakdown of LNA.

The net synergy profiles are constrained in the negative domain and 
show a hyperbolic trend, as a function of $\mu_{s}$. As $\mu_{s}$
increases, the negative domain gradually shrinks since synergy 
overpowers redundancy due to $\Delta I$ = $I_S - I_R$. In other words, 
the effective redundancy $I_{ER}$ decreases and gradually makes way 
for information independence. As conceptualised within the framework 
of PID, one can conveniently assume that redundancy takes care of 
common information sharing among the nodes of the diamond motif. 
Some of the key observations from the net synergy profiles shown in
Fig.~\ref{fig2}a, are as follows: For low $\mu_{s}$, 
\begin{eqnarray*}
\Delta I_{IM-DM}^{Add} > \Delta I_{BM-DM} > \Delta I_{IM-DM}^{Mult}.
\end{eqnarray*}

\noindent Similarly, for high $\mu_{s}$,  
\begin{eqnarray*}
\Delta I_{BM-DM} > \Delta I_{IM-DM}^{Add} > \Delta I_{IM-DM}^{Mult}.
\end{eqnarray*}

\noindent These trends can be justified as one takes into account the 
corresponding MI profiles shown in Fig.~\ref{fig2}b where the following 
ordering exists,
\begin{eqnarray*}
I_{IM-DM}^{Mult}(z;x,y) > I_{IM-DM}^{Mult}(z;x) = I_{IM-DM}^{Mult}(z;y) >
\nonumber \\
I_{IM-DM}^{Add}(z;x,y) > I_{IM-DM}^{Add}(z;x) = I_{IM-DM}^{Add}(z;y).
\end{eqnarray*}

\noindent In addition, one should also take note of the following inequality 
in the MI differences, 
\begin{eqnarray*}
I_{IM-DM}^{Mult}(z;x,y) - I_{IM-DM}^{Mult}(z;x) >  I_{IM-DM}^{Add}(z;x,y) - I_{IM-DM}^{Add}(z;x).
\end{eqnarray*}

\noindent Along with $I_{IM-DM}^{Mult}(z;y)$ and $I_{IM-DM}^{Add}(z;y)$, these differences 
ultimately guide the net synergy.

\noindent Furthermore, for low $\mu_{s}$,
\begin{eqnarray*}
I_{BM-DM}(s;x,y) > I_{BM-DM}(s;x) = I_{BM-DM}(s;y) > 
\nonumber \\
I_{IM-DM}^{Add}(z;x,y) > I_{IM-DM}^{Add}(z;x) = I_{IM-DM}^{Add}(z;y).
\end{eqnarray*}

\noindent On a similar note, for high $\mu_{s}$,
\begin{eqnarray*}
I_{IM-DM}^{Add}(z;x,y) > I_{IM-DM}^{Add}(z;x) = I_{IM-DM}^{Add}(z;y) >
\nonumber \\
I_{BM-DM}(s;x,y) > I_{BM-DM}(s;x) = I_{BM-DM}(s;y).
\end{eqnarray*}

\noindent We also note that $I_{BM-DM}(s;x,y)-I_{BM-DM}(s;x)$ 
decreases rapidly in comparison with 
$I_{IM-DM}^{Add}(z;x,y)-I_{IM-DM}^{Add}(z;x)$, 
whereas the MI profiles of BM-DM also decrease rapidly compared 
to MI profiles for additive and multiplicative IM-DM (see Fig.~\ref{fig2}b). 
This contrast in slowing down gets reflected in the net synergy. It is observed that 
$\Delta I_{BM-DM}$ occupies a long range and gets nearer to zero value 
in comparison with both $\Delta I_{IM-DM}^{Add}$ and 
$\Delta I_{IM-DM}^{Mult}$.

We also performed a parallel investigation on SNR, defined as 
$\Sigma^ {2} (s,z)/[\Sigma(s)\Sigma(z)-\Sigma^ {2} (s,z)]$ \cite{Bowsher2013}. 
SNR is calculated taking S as the signal and Z as the response of the DM. 
Both SNR$^{Add}$ and SNR$^{Mult}$ shows opposite trend with respect to 
$\Delta I_{IM-DM}^{Add}$ and $\Delta I_{IM-DM}^{Mult}$, respectively. We 
note that SNR$^{Mult}$ $>$  SNR$^{Add}$ and shows opposite trend of
$\Delta I_{IM-DM}^{Add} > \Delta I_{IM-DM}^{Mult}$. These trends are 
suggestive of the fact that with enhanced effective redundancy $I_{ER}$, 
the fidelity of the signaling pathway also gets strengthened. A similar 
observation was previously reported for a two-step cascade motif 
\cite{Biswas2016}. The observed agreement between the analytical and 
numerical results are indicative of the effectiveness of LNA applied in our 
calculation.

Here, we want to draw attention to the apparent similarities between 
net synergy based results from two different motifs namely TSC as in 
Ref.~\onlinecite{Biswas2016} and, for DM in the present manuscript.
Both systems show an ample amount of effective redundancy $I_{ER}$ 
and also the signal fidelity increases with it. It should be noted that
DM is constituted from TSC by duplicating its middle node and associated 
edges. For our current purpose, we carved out BM and IM from DM. 
It is important to point out that effective redundancy $I_{ER}$ gets 
manifested in TSC and DM using visibly different inter-relationship of the 
associated two-variable and three-variable MI-s. For a TSC motif 
S $\rightarrow$ X $\rightarrow$ Y, we got $I(s;x,y) \approx I(s;x) > I(s;y)$ 
both for the linear and nonlinear form of inter-species interactions \cite{Biswas2016}. 
Therefore, the net synergy picks up its major share from $I(s;y)$, i.e. 
$\Delta I(s;x,y) \approx -I(s;y)$. However, this is not the case in BM (equivalent to BM-DM) and,
 IM-DM (see Fig.~\ref{fig2}(a-b) and Fig.~\ref{fig5}(b) with Eq.~(\ref{eq7}-\ref{eq8})). In these architectures, none of 
 the individual three-variable and two-variable MI terms cancels each other 
 as in TSC case. All of these terms contribute to generate the net synergy profiles. 
 This trend also holds for IM which produces effective synergy $I_{ES}$  (see Fig.~\ref{fig5}(a) with Eq.~(\ref{eq8})).
 As an extra piece of observation related to this DM-centric study, we know that  
 effective redundancy $I_{ER}$ empowers signal fidelity with respect to both 
 additive and multiplicative signal integration mechanism. A TSC motif due to its 
 linear architecture, can not provide this specific insight. Hence, the creation of 
 effective redundancy $I_{ER}$ is mechanistically different in a TSC motif and 
 BM (BM-DM) and, IM-DM.


\begin{table}
\caption{
Table of the chemical reactions and associated propensities for the
diamond motif. Here, S, X, Y and, Z stand for biochemical species 
and $s$, $x$, $y$ and, $z$ represent copy numbers of the respective 
species expressed in molecules/$V$ with $V$ being the unit effective 
cellular volume. For X and Y mediated production of Z, an additive 
and a multiplicative logic is used. It results in two different synthesis 
rates, i.e. $k_z$ and $k_z'$, respectively. We have taken Hill coefficient 
$n=1$. The rate constants are expressed in min$^{-1}$.
}
\begin{ruledtabular}
\begin{tabular}{lll}
Biochemical  & Reaction & Propensity \\
Processes \\
\hline
Synthesis of S  & $\phi \rightarrow$ S & $k_s$ \\
Degradation of S  & S $\rightarrow \phi$ & $\mu_s s$ \\
S mediated \\
synthesis of X  & S $\rightarrow$ S + X & $k_x \frac{s^n}{K_1^n+s^n}$ \\
Degradation of X  & X $\rightarrow \phi$ & $\mu_x x$ \\
S mediated \\
synthesis of Y  & S $\rightarrow$ S + Y & $k_y \frac{s^n}{K_2^n+s^n}$ \\
Degradation of Y  & Y $\rightarrow \phi$ & $\mu_y y$ \\
X, Y mediated \\
synthesis of Z (Additive)           & X+Y $\rightarrow$ X+Y+Z & 
$k_z (\frac{x^n}{K_3^n+x^n}+\frac{y^n}{K_4^n+y^n})$\\
X, Y mediated \\
synthesis of Z (Multiplicative)           & X+Y $\rightarrow$ X+Y+Z & 
$k_z' \frac{x^n}{K_3^n+x^n}\frac{y^n}{K_4^n+y^n}$\\
Degradation of Z  & Z $\rightarrow \phi$ & $\mu_z z$ \\
\end{tabular}
\end{ruledtabular}
\end{table}


\begin{figure}[!t]
\begin{center}
\includegraphics[width=1.0\columnwidth,angle=0]{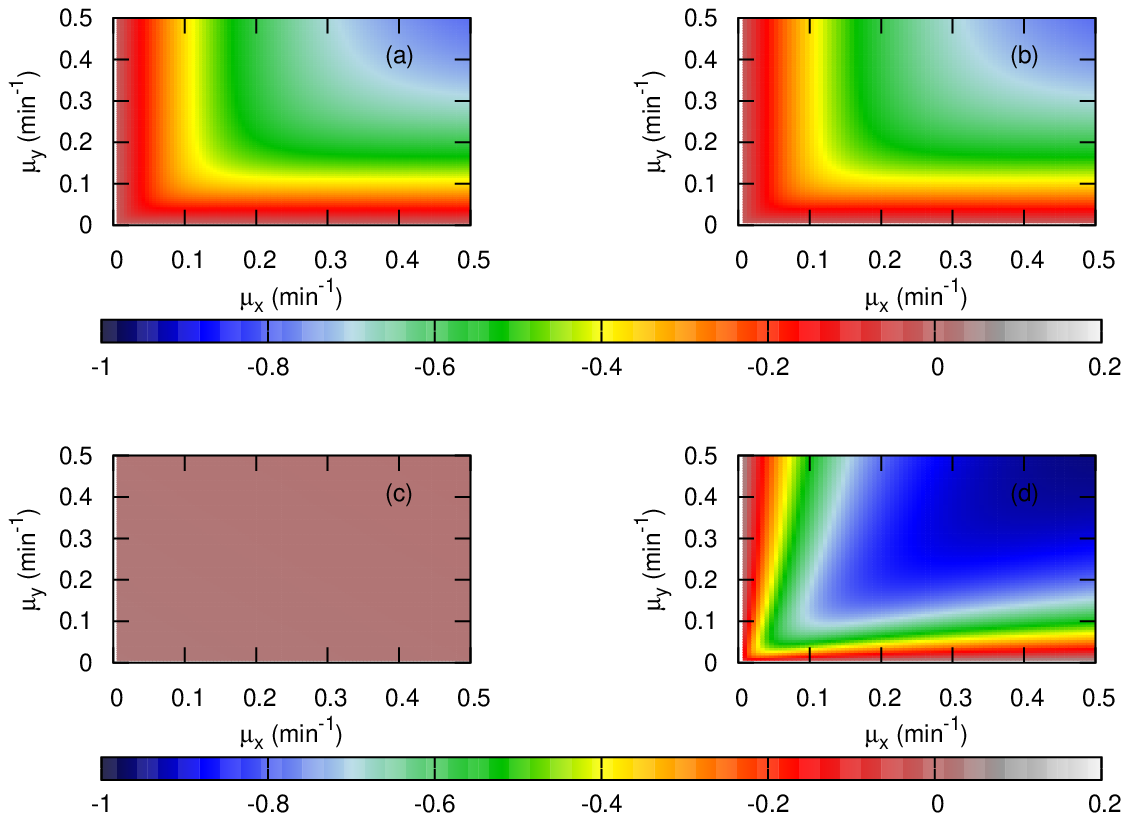}
\end{center}
\caption{Theoretical profiles (a)-(b) show variations in $\Delta I_{BM}(s;x,y)$ and $\Delta I_{BM-DM}(s;x,y)$ measured in bits as functions of $\mu_{x}$ and $\mu_{y}$, respectively. Similarly, (c)-(d) show variations in $\Delta I_{IM}^{Mult}(z;x,y)$ and $\Delta I_{IM-DM}^{Mult}(z;x,y)$ measured in bits as functions of $\mu_{x}$ and $\mu_{y}$, respectively. These maps are generated keeping $\mu_{s} =  0.1$ min$^{-1}$ and $\mu_{z} = 5$ min$^{-1}$. The synthesis rate parameters for BM are $k_s = \mu_s \langle s \rangle$, 
$k_x$ = $\mu_x \langle x \rangle ((K_1^n + \langle s \rangle^n)/\langle s \rangle^n)$ and, 
$k_y$ = $\mu_y \langle y \rangle ((K_2^n + \langle s \rangle^n)/\langle s \rangle^n)$. 
For IM the synthesis rates are
$k_x$ = $\mu_x \langle x \rangle$, 
$k_y$ = $\mu_y \langle y \rangle$ and, 
$k_z'$ = $\mu_z \langle z \rangle 
((K_3^n+\langle x \rangle^n)/\langle x \rangle^n) 
((K_4^n+\langle y \rangle^n)/\langle y \rangle^n)$. 
The synthesis rate parameters associated with DM are $k_s$ = $\mu_s \langle s \rangle$, 
$k_x$ = $\mu_x \langle x \rangle ((K_1^n+\langle s \rangle^n)/\langle s \rangle^n)$,
$k_y$ = $\mu_y \langle y \rangle ((K_2^n+\langle s \rangle^n)/\langle s \rangle^n)$ and, 
$k_z'$ = $\mu_z \langle z \rangle 
((K_3^n+\langle x \rangle^n)/\langle x \rangle^n)
((K_4^n+\langle y \rangle^n)/\langle y \rangle^n)$. We maintain $\langle s \rangle=10$, $\langle x \rangle=100$, $\langle y \rangle =100$, $\langle z \rangle=100$, $K_1=K_2=90$, $K_3=K_4=100$ all in molecules/$V$ with $V$ being the unit effective cellular volume and, $n=1$. 
}
\label{fig3}
\end{figure}


\begin{figure}[!t]
\begin{center}
\includegraphics[width=1.0\columnwidth,angle=0]{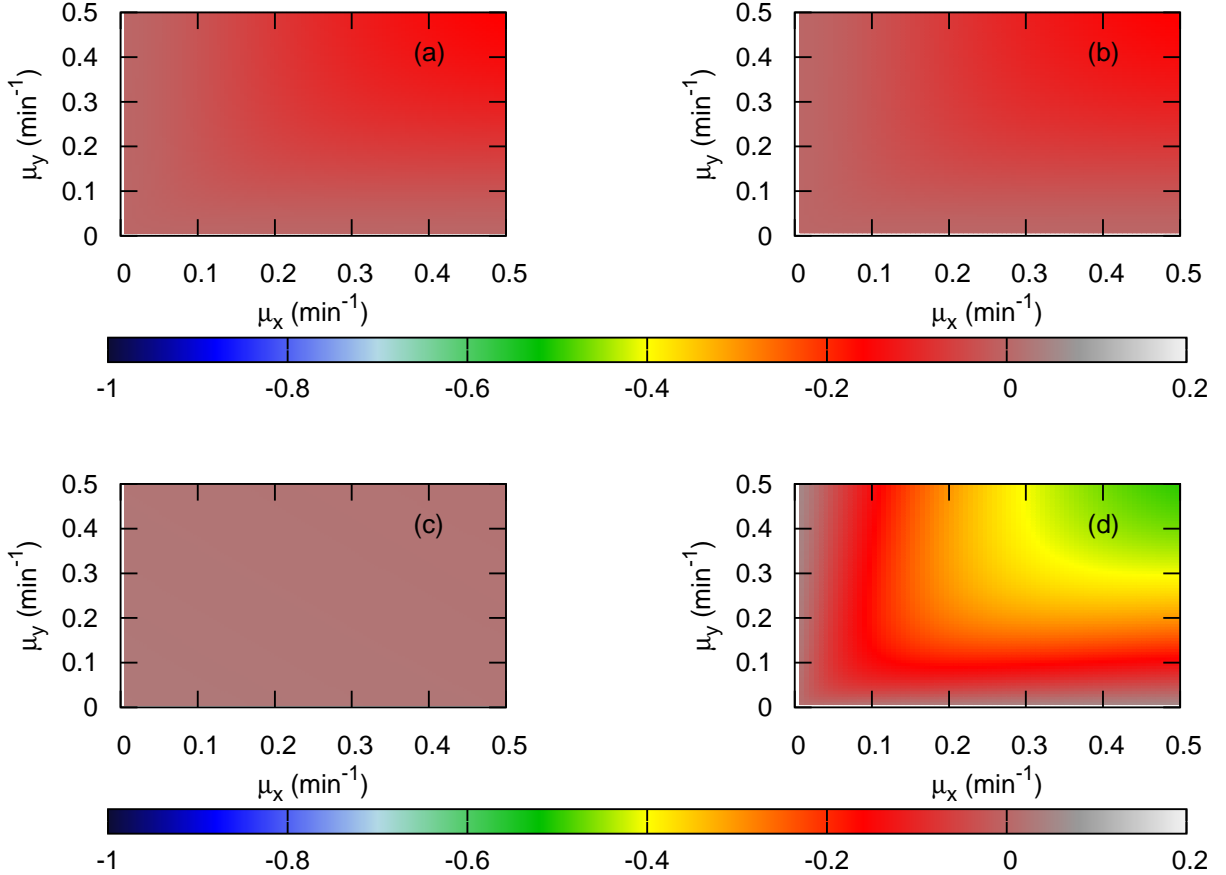}
\end{center}
\caption{Theoretical profiles (a)-(b) show variations in $\Delta I_{BM}(s;x,y)$ and $\Delta I_{BM-DM}(s;x,y)$ measured in bits as functions of $\mu_{x}$ and $\mu_{y}$, respectively. Similarly, (c)-(d) show variations in $\Delta I_{IM}^{Mult}(z;x,y)$ and $\Delta I_{IM-DM}^{Mult}(z;x,y)$ measured in bits as functions of $\mu_{x}$ and $\mu_{y}$, respectively. These maps are generated keeping $\mu_{s}=1$ min$^{-1}$ and $\mu_{z}= 5$ min$^{-1}$. The synthesis rate parameters for BM are as follows: $k_s = \mu_s \langle s \rangle$, 
$k_x$ = $\mu_x \langle x \rangle ((K_1^n + \langle s \rangle^n)/\langle s \rangle^n)$ and, 
$k_y$ = $\mu_y \langle y \rangle ((K_2^n + \langle s \rangle^n)/\langle s \rangle^n)$. 
For IM these are
$k_x$ = $\mu_x \langle x \rangle$, 
$k_y$ = $\mu_y \langle y \rangle$ and, 
$k_z'$ = $\mu_z \langle z \rangle 
((K_3^n+\langle x \rangle^n)/\langle x \rangle^n) 
((K_4^n+\langle y \rangle^n)/\langle y \rangle^n)$. 
The corresponding set for DM is as follows: $k_s$ = $\mu_s \langle s \rangle$, 
$k_x$ = $\mu_x \langle x \rangle ((K_1^n+\langle s \rangle^n)/\langle s \rangle^n)$, 
$k_y$ = $\mu_y \langle y \rangle ((K_2^n+\langle s \rangle^n)/\langle s \rangle^n)$ and, 
$k_z'$ = $\mu_z \langle z \rangle 
((K_3^n+\langle x \rangle^n)/\langle x \rangle^n)
((K_4^n+\langle y \rangle^n)/\langle y \rangle^n)$. We maintain $\langle s \rangle=10$, $\langle x \rangle=100$, $\langle y \rangle =100$, $\langle z \rangle=100$, $K_1=K_2=90$, $K_3=K_4=100$ all in molecules/$V$ with $V$ being the unit effective cellular volume and, $n=1$. 
}
\label{fig4}
\end{figure}


\subsection{Common source of input fluctuations generates effective redundancy}

The two maps in Fig.~\ref{fig3}(a,b) are generated by scanning 
the parameter spaces of $\mu_{x}$ and $\mu_{y}$ for fixed signal 
relaxation rate $\mu_{s} = 0.1$ min$^{-1}$ and depict the net synergy 
$\Delta I_{BM} (s;x,y)$ and $\Delta I_{BM-DM} (s;x,y)$ of BM and 
BM-DM, respectively. Fig.~\ref{fig4}(a,b) do the same for 
$\mu_{s} = 1$ min$^{-1}$. The maps of the net synergy 
$\Delta I_{IM}^{Mult} (z;x,y)$ and $\Delta I_{IM-DM}^{Mult} (z;x,y)$ 
of IM and IM-DM, respectively, are also shown in Fig.~\ref{fig3}(c,d) 
and Fig.~\ref{fig4}(c,d) for  $\mu_s = 0.1$ min$^{-1}$ and 
$\mu_{s} = 1$ min$^{-1}$, respectively. 
Fig.~\ref{fig3}(c) shows positive net synergy $\Delta I_{IM}^{Mult} (z;x,y)$ 
or effective synergy $I_{ES}$ spanning the entire parameter range. On the 
other hand, Fig.~\ref{fig3}(d) shows redundancy is empowered in comparison 
with synergy in most of the area of the net synergy profile 
$\Delta I_{IM-DM}^{Mult} (z;x,y)$ consisting of a larger negative domain along 
with a minuscule positive valued region. In short, moving from Fig.~\ref{fig3}(c) 
to Fig.~\ref{fig3}(d), effective redundancy $I_{ER}$ takes up control over 
effective synergy $I_{ES}$.  
In comparison with their corresponding counterparts in Fig.~\ref{fig3}, 
Fig.~\ref{fig4}(a,d) shows the interplay between synergy and redundancy to 
be less favorable to redundant information. This is reflected in the 
shrunken range of variation of the net synergy.

To comprehend the nature of the net synergy profiles, we take note of 
the fact that there are multiple time scales involved with the biochemical 
species constituting different topologies (BM, BM-DM, IM, IM-DM and, DM). 
These time scales play crucial roles to affect the information flow along the 
motif. They sometimes facilitate propagation of information and hinder 
otherwise \cite{Ronde2012,Bruggeman2009,Hinczewski2014,Maity2015}. 
Now, we re-express the preceding statement as a guiding principle to 
analyze information flow in the motif under investigation.
Whenever the upstream species concentration fluctuates slowly as 
compared to its immediate downstream species (i.e. the upstream 
species has got a relatively small relaxation rate with respect to that 
of the downstream species) it helps the downstream species to sense 
the upstream fluctuations accurately thereby allowing information flow 
to occur. In the opposite scenario, the downstream species fails to 
follow the rapid upstream fluctuations with adequate precision thereby 
obstructing the information propagation. In both the cases mentioned, 
the corresponding effects get pronounced depending upon the extent of 
separation of these time scales with respect to each other 
\cite{Maity2015}. The idea of separation of time scales seems very useful 
to rationalize this low-pass filter like characteristics of DM. Surprisingly, 
this simple physical analog sets up the logical framework which 
encapsulates the phenomenon of frequency dependent encoding, 
processing and, decoding mechanisms of biological signals 
\cite{Hansen2015}.

The net synergy profiles in Fig.~\ref{fig2}(a) can be well understood 
keeping the above-mentioned principle in perspective. The range of 
variation of $\mu_{s}$ spans regions with $\mu_{s} < \mu_{x} (\mu_{y})$, 
$\mu_{s} = \mu_{x} (\mu_{y})$ and $\mu_{s}>\mu_{x} (\mu_{y})$. At the 
same time, by keeping $\mu_{z} = 5$ min$^{-1}$ which is 10 times faster 
than $\mu_{x} (\mu_{y})$, fixed at $0.5$ min $^{-1}$; adequate amount 
of information flow is allowed for convenience. It is clear from 
Fig.~\ref{fig2}(a) that as the source species fluctuates faster compared 
to downstream species, transmitted amount of effective redundancy 
$I_{ER}$ decreases.

In Fig.~\ref{fig3}(a), as we move along the diagonal from low 
$\mu_{x} (\mu_{y}$) to high $\mu_{x} (\mu_{y}$), it is observed that 
the value of net synergy $\Delta I_{BM} (s;x,y)$ becomes more 
negative implying increase in effective redundancy $I_{ER}$. 
Negative $\Delta I$ means $I_R>I_S$ and increment in negativity 
may be caused either keeping $I_S$ fixed and increasing $I_R$ or 
changing both $I_S$ and $I_R$ while maintaining the rate of change of 
$I_R$ $>$ the rate of change of $I_S$. In this sense, we infer 
that $I_{ER}$ rises. It should be noted that moving along this 
diagonal direction, both the downstream species X and Y become 
more sensitive towards the signal (S) fluctuations thereby get to 
harness more information. By this token, according to the definitions 
of $\Delta I$ prescribed by PID, the amount of common or redundant 
information content between X and Y about S increases causing the 
net synergy to decrease.

One can compare the net synergy profiles of BM in Figs.~\ref{fig3}(a) 
and \ref{fig4}(a) subject to variation in the signal relaxation. It is 
visible that with increased $\mu_{s}$ from 0.1 min$^{-1}$ to 1 min$^{-1}$, 
the domain of net synergy shifts close to zero value. Such a change in the 
nature of the net synergy profile is due to the fact that in Fig.~\ref{fig4}(a) 
fixing $\mu_{s}=1$ min$^{-1}$ blocks information flow significantly in the 
motif since both the downstream species X and Y fluctuates in the range 
of $0 - 0.5$ min$^{-1}$, slower than the time scale of fluctuations of the 
source species S.

In Fig.~\ref{fig3}(c), the net synergy $\Delta I_{IM}^{Mult} (z;x,y)$ is 
entirely constrained in the positive domain, i.e. we get effective synergy 
$I_{ES}$. Since we have not specified synergy and redundancy 
independently in a quantitative manner, we are not certain that whether 
this positive nature is due to pure synergy or synergy being dominant 
over redundancy, but we can at least make an intuitive inference. It has 
a certain benefit not to opt for any specific definition of any of the PID 
terms, thereby keeping ample generality in the treatment of the motif 
which may be subjected to some hitherto unknown  biological 
constraints. Hence, any concrete comment on the extent of both 
synergy and redundancy can be made only in a comparative manner 
using a relative scaling. Redundancy being sharing of information 
\cite{Barrett2015}, may originate from a common source which in this 
case (i.e. IM) is absent. There are two uncorrelated source X and Y 
and single target Z in IM unlike in BM where the targets X and Y share 
a common source S. We conjecture here that the absence of a common 
source is suggestive of pure synergy over the entire parameter space 
(Figs.~\ref{fig3}(c) and \ref{fig4}(c)). 
At par with same line of argument,
it can be stated that, effective redundancy $I_{ER}$ 
 arises in TSC motif (S $\rightarrow$ X $\rightarrow$ Y), 
 because information source variable S is common for the information target variables X and Y while computing
 the net synergy $\Delta I(s;x,y)$ in Ref~\onlinecite{Biswas2016}. 
 In the dynamical sense also, the effect of signaling source S is 
 stochastically manifested in X first and finally in Y via X.
 Therefore, the concept of commonality in input fluctuations creating 
 effective redundancy $I_{ER}$ remains in place. Information-theoretic analysis
 grounded in neurophysiological recordings in the primary visual context of anesthetized macaque 
 monkeys has successfully applied the information breakdown methodology (IBM) \cite{Montani2007}, 
 similar to PID discussed in our current work. We invoke this particular analysis to place
 some intuitive aspects of redundant and synergistic information in our model system
 through a qualitative mapping from neuronal network to GRN. As in Ref.~\onlinecite{Montani2007,Montangie2016}, 
 IBM prescribes the following decomposition:
 
 \begin{equation}
 \label{eq15}
 I_{ensemble}=I_{lin}+I_{sigsim}+I_{cor,dep}+I_{cor,ind}.
 \end{equation}

\noindent
For BM and BM-DM, $I_{ensemble}^{BM (BM-DM)}(s;x,y) \equiv I_{BM (BM-DM)}(s;x,y)$, $I_{lin}^{BM (BM-DM)} \equiv I_{BM (BM-DM)}(s;x)+ \\
I_{BM (BM-DM)}(s;y)$. Here, we denote both $I_{ensemble}^{BM}(s;x,y)$ and $I_{ensemble}^{BM-DM}(s;x,y)$ by a single notation $I_{ensemble}^{BM (BM-DM)}(s;x,y)$ and we use this type of notation for the rest of the information terms considering every motifs and sub-motifs.    
Therefore, $I_{ensemble}^{BM (BM-DM)}(s;x,y)-I_{lin}^{BM (BM-DM)}=I_{BM (BM-DM)}(s;x,y)-I_{BM (BM-DM)}(s;x) \\
- I_{BM (BM-DM)}(s;y)$. According to PID, the expression on the right-hand side
of the last expression is defined as the net synergy, $\Delta I=I_{S}-I_{R}$ (see Eq.~\ref{eq9}). Taking the lead from Ref~\onlinecite{Montani2007}, we put $I_{sigsim}^{BM (BM-DM)}(s;x,y)$ into perspective of GRN. This signal similarity measure accounts for the  reduction in $I_{ensemble}^{BM (BM-DM)}(s;x,y)$ caused by the similar type of activation mechanisms represented by the parameters associated with the edges, e.g. S $\rightarrow$ X and S $\rightarrow$ Y
in BM and BM-DM. The corresponding production terms (See Table~I) are similar as $K_{1}=K_{2}$, $k_{x}=k_{y}$
(given $\langle x \rangle = \langle y \rangle$ and $\mu_{x}=\mu_{y}$) and $n=1$ for both species X and Y. Thus, this component
contributes negatively for BM and, BM-DM. $I_{cor,dep}$ and $I_{cor,ind}$ designates signal-dependent correlation induced information 
and signal-independent correlation induced information, respectively for all motifs and sub-motifs. By information-theoretic definition, $I_{cor,dep} \ge 0$ in general. Since in our model system, all the correlation functions are non-negative, and noise profiles are uncorrelated (see section II for the noise characteristics),
$I_{cor,ind}$ should be zero for all the motifs and sub-motifs. In the case of BM and BM-DM, if we take into consideration the presence of a  common source of fluctuations (S), we can justify the predominantly negative trend in their net synergy profiles by utilizing the negative contribution of $I_{sigsim}^{BM (BM-DM)}(s;x,y)$. Considering these points, we re-write Eq.~\ref{eq15} for BM and BM-DM as follows:

\begin{equation}
\Delta I_{BM (BM-DM)}(s;x,y)=|I_{cor,dep}^{BM (BM-DM)}(s;x,y)|-|I_{sigsim}^{BM (BM-DM)}(s;x,y)|.
\end{equation}

\noindent
Where $|...|$ designate the magnitude of corresponding information component. For IM and IM-DM, $I_{ensemble}^{IM (IM-DM)}(z;x,y) \equiv I_{IM (IM-DM)}(z;x,y)$, $I_{lin}^{IM (IM-DM)} \equiv I_{IM (IM-DM)}(z;x)+I_{IM (IM-DM)}(z;y)$.
Therefore, $I_{ensemble}^{IM (IM-DM)}-I_{lin}^{IM (IM-DM)}=I_{IM (IM-DM)}(z;x,y)-I_{IM (IM-DM)}(z;x)-I_{IM (IM-DM)}(z;y)$. These expressions are valid for both additive and multiplicative signal integration mechanisms. We will further denote additive and multiplicative integration scheme by writing \lq$-Add $\rq , \lq$-Mult$\rq ~respectively. Now, for the edges X $\rightarrow$ Z and Y $\rightarrow$ Z
in IM-Add, IM-Mult, IM-DM-Add and, IM-DM-Mult, we have architectural similarity maintained by similar production functions with a single $k_{z}$, $k_{z}'$, $K_{3}=K_{4}$, $k_{x}=k_{y}$ (given $\langle x \rangle = \langle y \rangle$ and $\mu_{x}=\mu_{y}$) and $n=1$. However, surprisingly for IM-Add and IM-Mult, even with $\Sigma(x,y)=0$ (implying $I_{cor,dep}^{IM-Add}(z;x,y)=I_{cor,dep}^{IM-Mult}(z;x,y)=0$), 
we have $\Delta I_{IM-Add (IM-Mult)}(z;x,y)>0$. To rationalize this observation, we note that since in IM-Add and IM-Mult, there is no common source of fluctuations present but the system still possesses similar signaling branches, $I_{sigsim}$ contributes positively as follows:

\begin{equation}
\Delta I_{IM-Add (IM-Mult)}(z;x,y)=|I_{sigsim}^{IM-Add (IM-Mult)}(z;x,y)|.
\end{equation}

\noindent
In IM-DM-Add and IM-DM-Mult, due to the presence of common input S and similarities in between two bifurcating branch (S $\rightarrow$ X and S $\rightarrow$ Y) and similarities in between two integrating branch (X $\rightarrow$ Z and Y $\rightarrow$ Z) $I_{sigsim}^{IM-DM-Add}(z;x,y)$ and $I_{sigsim}^{IM-DM-Mult}(z;x,y)$ is negative as in BM and BM-DM. And by and large it overpowers $I_{cor,dep}^{IM-DM-Add}(z;x,y)$ and $I_{cor,dep}^{IM-DM-Mult}(z;x,y)$ respectively to make $\Delta I_{IM-DM-Add}(z;x,y)<0$ and $\Delta I_{IM-DM-Mult}(z;x,y)<0$ respectively according to the following two consecutive expressions:

\begin{equation}
\Delta I_{IM-DM-Add}(z;x,y)=|I_{cor,dep}^{IM-DM-Add}(z;x,y)|-|I_{sigsim}^{IM-DM-Add}(z;x,y)|.
\end{equation}

\begin{equation}
\Delta I_{IM-DM-Mult}(z;x,y)=|I_{cor,dep}^{IM-DM-Mult}(z;x,y)|-|I_{sigsim}^{IM-DM-Mult}(z;x,y)|.
\end{equation}

\noindent
This rudimentary connection between PID and IBM is far from complete in the present model GRN systems and at best offers some speculative remarks.
 
 In Figs.~\ref{fig3}(d) and \ref{fig4}(d), the net synergy decreases along the 
 diagonal from low $\mu_{x} (\mu_{y}$) to high $\mu_{x} (\mu_{y}$) which is at par with 
the previously placed argument based on the idea of separation of time 
scales. In Ref.~\onlinecite{Barrett2015}, a dynamical system analog of 
IM was showcased and was noted that synergy between sources about 
the target, is a function of the correlation between sources. Although, there 
is no physical link in between the sources, (cross) correlation has been 
invoked in their noise processes. Our treatment does not involve noise 
correlation and for that matter, depends on inter-species interactions to 
develop correlations that further dictate the net synergy. This is rather 
straight-forward and does not create any misunderstanding when one 
tries to interpret information processing based on the connectivity 
diagrams of the motifs. The specialized PID namely MMI PID 
\cite{Barrett2015} defines redundancy as the minimum MI provided by 
the sources about the target. Synergy comes from the extra information 
by the weaker source given the other source is known. However, in more 
generalized situations, the motifs can have architecturally redundant 
pathways, i.e. source variables may be equally strong. This is one of 
the key observations from our study where we set out to map 
architectural redundancy and informational redundancy onto each other.  


\begin{figure}[!t]
\begin{center}
\includegraphics[width=1.0\columnwidth,angle=0]{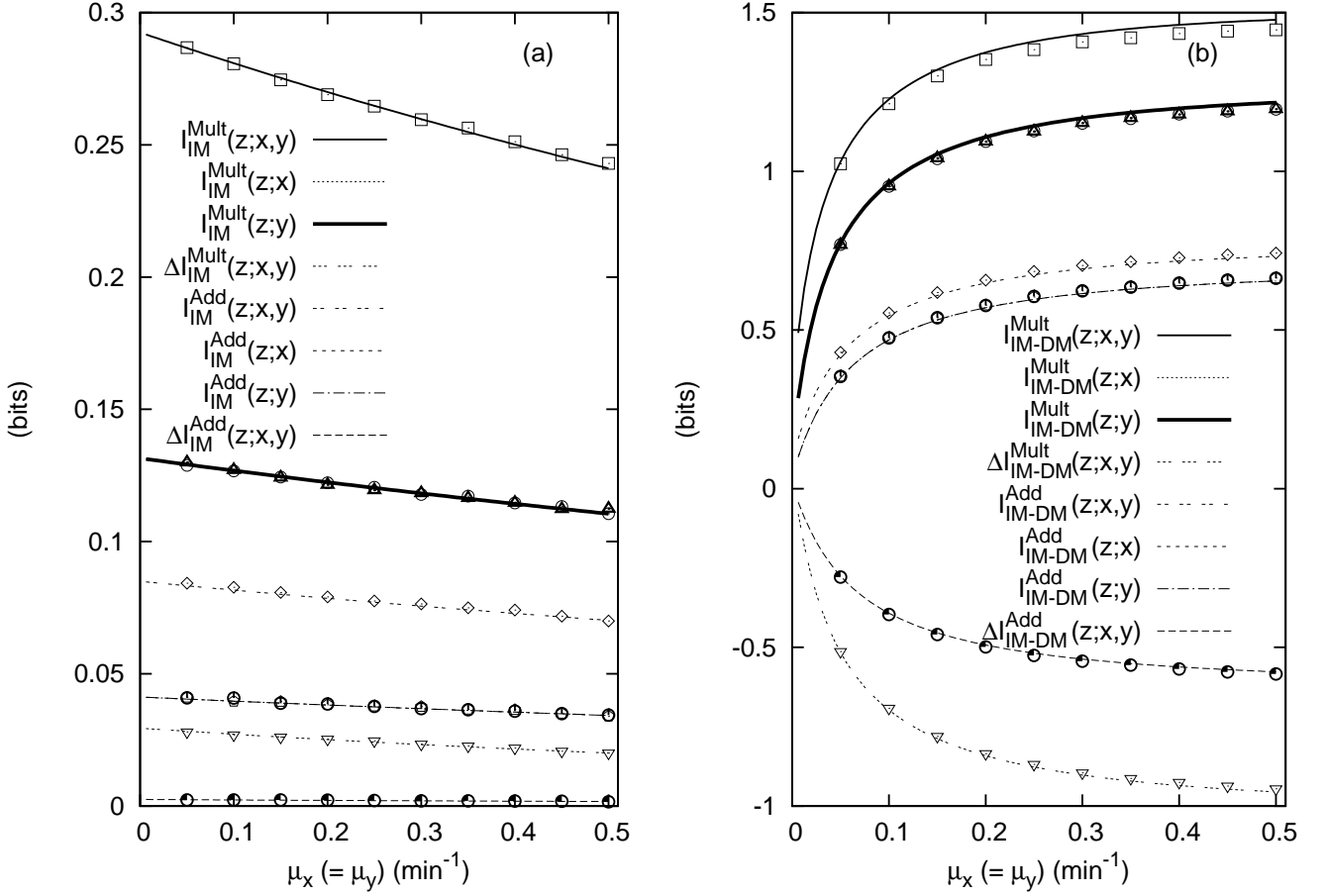}
\end{center}
\caption{Individual MI and the net synergy profiles for (a) IM and, (b) IM-DM as functions of $\mu_{x}(=\mu_{y})$. The lines are due to analytical results and the symbols represent numerical data generated using Gillespie's algorithm \cite{Gillespie1976,Gillespie1977} with an  ensemble averaging over $10^6$ independent time series. These profiles are generated keeping $\mu_{s}=0.1$ min$^{-1}$ and $\mu_{z}= 5$ min$^{-1}$. 
For IM the synthesis rate parameters are
$k_x$ = $\mu_x \langle x \rangle$, 
$k_y$ = $\mu_y \langle y \rangle$,
$k_z$ = $\mu_z \langle z \rangle (\frac{\langle x \rangle^n}{K_3^n+\langle x \rangle^n}+\frac{\langle y \rangle^n}{K_4^n+\langle y \rangle^n})^{-1}$ and, 
$k_z'$ = $\mu_z \langle z \rangle 
((K_3^n+\langle x \rangle^n)/\langle x \rangle^n) 
((K_4^n+\langle y \rangle^n)/\langle y \rangle^n)$. 
The corresponding set for DM is as follows: 
$k_s$ = $\mu_s \langle s \rangle$, 
$k_x$ = $\mu_x \langle x \rangle ((K_1^n+\langle s \rangle^n)/\langle s \rangle^n)$, 
$k_y$ = $\mu_y \langle y \rangle ((K_2^n+\langle s \rangle^n)/\langle s \rangle^n)$,
$k_z$ = $\mu_z \langle z \rangle (\frac{\langle x \rangle^n}{K_3^n+\langle x \rangle^n}+\frac{\langle y \rangle^n}{K_4^n+\langle y \rangle^n})^{-1}$ and, 
$k_z'$ = $\mu_z \langle z \rangle 
((K_3^n+\langle x \rangle^n)/\langle x \rangle^n)
((K_4^n+\langle y \rangle^n)/\langle y \rangle^n)$. We maintain $\langle s \rangle=10$, $\langle x \rangle=100$, $\langle y \rangle =100$, $\langle z \rangle=100$, $K_1=K_2=90$, $K_3=K_4=100$ all in molecules/$V$ with $V$ being the unit effective cellular volume and, $n=1$. 
 }
\label{fig5}
\end{figure} 

For a comparative study to explore how integration mechanism 
distinguishes IM and IM-DM, we set out to vary $\mu_{x}(=\mu_{y})$ 
keeping $\mu_{s}$ constant and we observe in Fig.~\ref{fig5}(a-b): 
\begin{eqnarray*}
\Delta I_{IM}^{Mult}(+) > \Delta I_{IM}^{Add}(+) > 
\Delta I_{IM-DM}^{Add}(-) > \Delta I_{IM-DM}^{Mult}(-). 
\end{eqnarray*}

\noindent By putting $+$  or  $-$ signs we denote the associated 
quantities to be positive or negative valued, respectively. To probe 
it further, we plotted the individual MI keeping the conditions unaltered 
in Fig.~\ref{fig5}(a-b) and notice the trend 
\begin{eqnarray*}
I_{IM-DM}^{Mult}(z;x,y) > I_{IM-DM}^{Mult}(z;x) = I_{IM-DM}^{Mult}(z;y) >
\nonumber \\
I_{IM-DM}^{Add}(z;x,y) > I_{IM-DM}^{Add}(z;x) = I_{IM-DM}^{Add}(z;y) >
\nonumber \\
I_{IM}^{Mult}(z;x,y) > I_{IM}^{Mult}(z;x) = I_{IM}^{Mult}(z;y) >
\nonumber \\
I_{IM}^{Add}(z;x,y) > I_{IM}^{Add}(z;x) = I_{IM}^{Add}(z;y).
\end{eqnarray*}


\subsection{Activation coefficients modulate effective redundancy 
and explain the trade-off between binding affinities of different 
biochemical species}

The four edges of DM are characterized by three types of parameters 
namely the activation coefficients ($K_1, K_2, K_3$ and, $K_4$), the 
synthesis rates of biochemical species  ($k_s, k_x, k_y$ and, $k_z$) 
and the Hill coefficient ($n$) of the input regulatory functions. 
This minimal set of parameters ($K_1, K_2, K_3, K_4, k_s, k_x, k_y, k_z$ 
and, $n$) can be tuned during an experiment. It has been argued that in 
a continuous changing environment, these parameters associated with 
the population of each of the biochemical species, encounter selection 
pressure and thus precisely optimize the expression levels \cite{Alon2006}. 
The time scales of their adaptability is typical of the order of hundred of 
generations. 
Experimentally, $K_i (i =1,2,3,4)$ can be altered by inducing mutations 
in the DNA sequences of the promoter region where the activator or repressor 
molecules bind. To alter $k_i$ $(i = s,x,y,z)$, the binding site 
sequence for RNA polymerase are mutated. If suitable parameterization 
associated with a motif is done with these numbers, then experiments 
can be designed to place the biological motifs under selection pressure. 
Uri Alon points out that the resulting changes that an organism 
acquires under such pressure are inheritable through generations facing 
diverse environmental conditions \cite{Alon2006}. These phenomenological 
inputs motivate us to link information processing with these numbers. 
In the previous Figs.~(\ref{fig2}-\ref{fig5}), we have already done so by 
varying $\mu_i$ $(i = s,x,y)$, thereby making variations in $k_i$ $(i = s,x,y)$ 
since the constraint of fixed population size at steady state makes 
$k_i \propto \mu_i$ $(i = s,x,y,z)$. We further reiterate that in our 
calculation we have used $n=1$ to retain the level of nonlinearity in the 
input regulatory functions analytically tractable under the purview of LNA. 


\begin{figure}[!t]
\begin{center}
\includegraphics[width=1.0\columnwidth,angle=0]{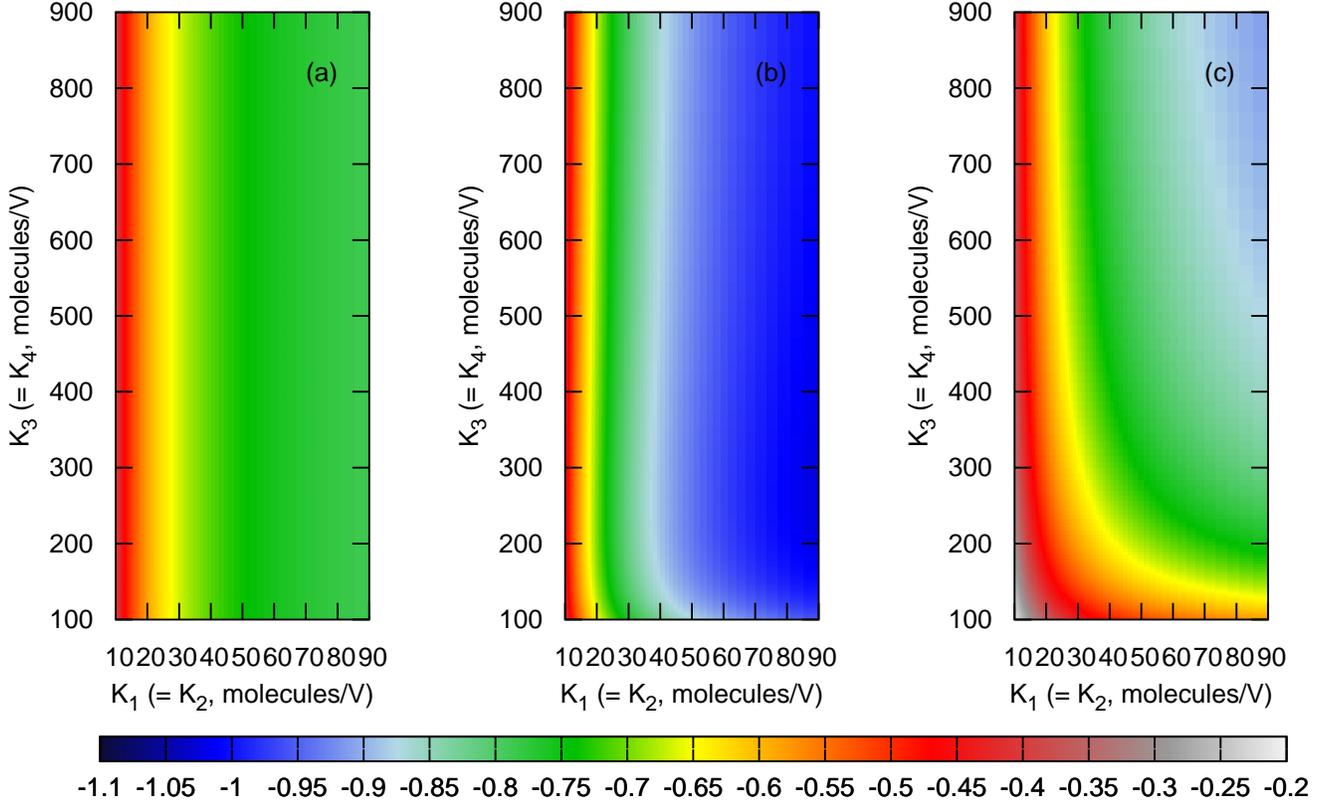}
\end{center}
\caption{The maps (a),(b) and, (c) portray $\Delta I_{BM-DM}(s;x,y)$, $\Delta I_{IM-DM}^{Mult}(z;x,y)$ and, $\Delta I_{IM-DM}^{Add}(z;x,y)$ as functions of $K_{1}(=K_{2})$ and $K_{3}(=K_{4})$ respectively, in the unit of bits.
The populations of the biochemical species S, X, Y and, Z are fixed at steady state  $\langle s \rangle=10$, $\langle x \rangle=100$, $\langle y \rangle=100$ and, $\langle z \rangle=100$ all in molecules/$V$ with $V$ being the unit effective cellular volume. The relevant rate parameters are $k_s$ = $\mu_s \langle s \rangle$, $k_x$ = $\mu_x \langle x \rangle ((K_1^n+\langle s \rangle^n)/\langle s \rangle^n)$, 
$k_y$ = $\mu_y \langle y \rangle ((K_2^n+\langle s \rangle^n)/\langle s \rangle^n)$, 
$k_z$ = $\mu_z \langle z \rangle (\frac{\langle x \rangle^n}{K_3^n+\langle x \rangle^n}+\frac{\langle y \rangle^n}{K_4^n+\langle y \rangle^n})^{-1}$ and, 
$k_z'$ = $\mu_z \langle z \rangle 
((K_3^n+\langle x \rangle^n)/\langle x \rangle^n)
((K_4^n+\langle y \rangle^n)/\langle y \rangle^n)$. We use $\mu_{s}=$ 0.1 min$^{-1}$, $\mu_{x}=\mu_{y}=$ 0.5 min$^{-1}$, $\mu_{z}=5$ min$^{-1}$ and, $n=1$ to generate the theoretical figures.   
}
\label{fig6}
\end{figure}

Figure~\ref{fig6}(a,b,c) portrays the effect of variations of $K_1$, 
$K_2$, $K_3$ and, $K_4$ on the net synergy 
$\Delta I_{BM-DM} (s;x,y)$, $\Delta I_{IM-DM}^{Mult} (z;x,y)$ and, 
$\Delta I_{IM-DM}^{Add} (z;x,y)$, respectively. We have kept 
$\mu_{s}=0.1$ min$^{-1}$, $\mu_{x}=\mu_{y}=0.5$ min$^{-1}$, 
$\mu_{z}=5$ min$^{-1}$ to show the above-mentioned variations 
subject to the constraint $K_1 = K_2$ and $K_3 = K_4$. By 
making such a set of arrangements, we have ensured adequate 
information flow in the motif while keeping architectural redundancy 
(based on interactions) in between both the bifurcating and the 
integrating branches. Figure~\ref{fig6}(a) indicates that with 
increasing $K_1 (K_2)$, redundancy contributes more, making 
the net synergy more negative. We can not  rule out the presence 
of synergy, but it is overpowered by redundancy resulting in the  
production of effective redundancy $I_{ER}$. Similarly in 
Fig.~\ref{fig6}(b), we show that with high $K_1 (K_2)$ and low 
$K_3 (K_4)$, the two integrating branches of the diamond motif, 
contribute the maximum level of redundancy in the entire map 
with $\Delta I_{IM-DM}^{Mult}(z;x,y)$ giving the most pronounced 
effects. In this regime, the intermediate information sources X and 
Y which are activated by the source S with low strength, in return, 
activate target Z with moderate strength. Fig.~\ref{fig6}(c) presents 
the profile of $\Delta I_{IM-DM}^{Add}(z;x,y)$ contributing maximum 
effective redundancy $I_{ER}$, where S interacts with X and Y 
weakly and so does both X and Y with Z. In this respect, we restate 
that the strength of activation is determined by the parameters 
$K_1, K_2, K_3$ and, $K_4$.   
To be specific, by using $K_1 = K_2 = 10-90$ and 
$K_3 = K_4 =100-900$ (all in molecules/$V$ with $V$ being the 
unit effective cellular volume) \cite {Buchler2009}, the nonlinear 
regulatory functions ($s^n/(K_1^n+s^n)$, $s^n/(K_2^n+s^n)$, 
$x^n/(K_3^n+x^n)$ and, $y^n/(K_4^n+y^n)$) in the production 
terms for X, Y and, Z, respectively, take numerical values in the 
range of 0.1 (weak activation) - 0.5 (moderate activation) at 
steady state.

The maps in Fig.~\ref{fig6}(a,b,c) show that, with decreasing 
strength of activation of the intermediate X and Y by S and by 
increasing the activation strength of Z by X and Y, the system 
acquires more redundancy for the BM-DM and the IM-DM-Mult 
as the case may be. For IM-DM-Add, the system chooses a  
decrease in the activation strength of Z by X and Y, to increase 
effective redundancy $I_{ER}$ while, still holding weak interaction 
of S with X and Y. Here we redirect our attention to Fig.~\ref{fig2}(a,b,c) 
where we have observed that SNR, the metric for fidelity in 
information processing, \cite{Bowsher2013} increases with 
increasing effective redundancy $I_{ER}$ . Keeping these points 
in mind, we propose that under weak activation levels of X and Y 
(i.e. low steady-state values of the nonlinear terms $s^n/(K_1^n+s^n)$, 
etc.) along with, either moderate activation levels of the target Z for 
multiplicative signal integration mechanism (i.e. moderate steady-state 
values of the nonlinear terms $x^n/(K_3^n+x^n)$, etc.) or 
weak activation levels of the target Z for additive signal integration 
mechanism (i.e. low steady-state values of the nonlinear terms 
$x^n/(K_3^n+x^n)$, etc.) maximum level of effective redundancy 
is likely to be selected naturally by incorporating suitable mutations. 
Therefore, noise minimizing biological motifs will eventually favor 
regulation of the above-mentioned strengths for biochemical species 
S, X, Y and, Z.

One can find a similar situation in case of regulatory DNA sequence 
motifs where for a single transcription factor, there exist more than 
one operators with varying affinity. This phenomenon has been 
ascribed to the evolutionary tuning of the binding affinities so that 
the regulated gene can be made maximally functional. Here, one 
should keep in mind that these sequences are constantly exposed 
to simultaneous forces of natural selection and mutation. In essence, 
this links the genotype which is acted upon by mutation and, the phenotype which is 
selected if deemed fit in the backdrop of environmental changes 
\cite{Gerland2002}. Our predictions find resonance in the fact that 
high sequence specificity may be harmful to the transcription control 
network in the face of deleterious mutation. The system also keeps a  
check on the weaker end of specificity spectrum to avoid spurious 
interactions. This trade-off essentially provides the required 
robustness in the control network \cite {Sengupta2002}.


\begin{figure}[!t]
\begin{center}
\includegraphics[width=1.0\columnwidth,angle=0]{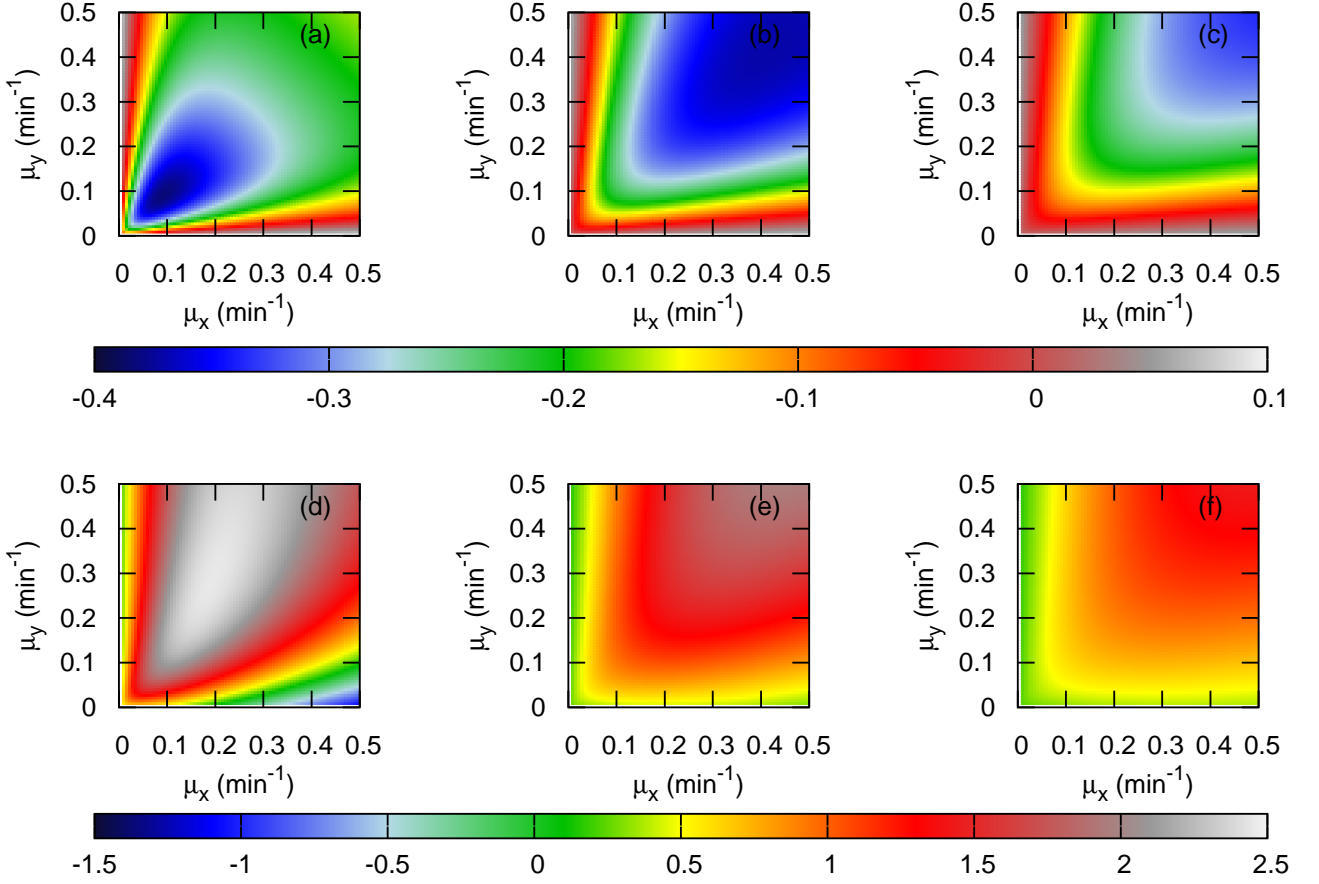}
\end{center}
\caption{Panels (a)-(c) depict $\Delta I_{IM-DM}^{Mult}(z;x,y)-\Delta I_{BM-DM}(s;x,y)$ (in bits) as functions of $\mu_{x}$ and $\mu_{y}$ for $\mu_{s}=$ 0.1, 0.5 and, 1 min$^{-1}$, respectively. Similarly, panels (d)-(f) show SNR$_{IM-DM}^{Mult}-$SNR$_{BM-DM}$ for $\mu_{s}=$ 0.1, 0.5 and, 1 min$^{-1}$, respectively. SNR$_{BM-DM}$ and SNR$_{IM-DM}^{Mult}$ are computed considering the signaling pathways S $\rightarrow$ X and X $\rightarrow$ Z, respectively. The theoretical expressions are obtained using constant populations (at steady state) of the biochemical species: $\langle s \rangle=10$, $\langle x \rangle=100$, $\langle y \rangle=100$ and, $\langle z \rangle=100$ all in molecules/$V$ with $V$ being the unit effective cellular volume. The expressions for the rate parameters are $k_s$ = $\mu_s \langle s \rangle$, $k_x$ = $\mu_x \langle x \rangle ((K_1^n+\langle s \rangle^n)/\langle s \rangle^n)$, 
$k_y$ = $\mu_y \langle y \rangle ((K_2^n+\langle s \rangle^n)/\langle s \rangle^n)$ and, 
$k_z'$ = $\mu_z \langle z \rangle 
((K_3^n+\langle x \rangle^n)/\langle x \rangle^n)
((K_4^n+\langle y \rangle^n)/\langle y \rangle^n)$. We have kept $K_1=K_2=90$, $K_3=K_4=100$ all in molecules/$V$ with $V$ being the unit effective cellular volume, $\mu_{z}=5$ min$^{-1}$ and, $n=1$ throughout to generate the theoretical profiles.   
}
\label{fig7}
\end{figure}


\subsection{Parametric dependance of net synergy contributions of 
different sub-motifs reveals the corresponding noise-tolerant sub-motif}

To capture how the bifurcation and integration sub-motifs perform 
relative to each other in terms of their individual net synergy 
contributions, we show 
$\Delta I_{IM-DM}^{Mult} (z;x,y) - \Delta I_{BM-DM} (s;x,y)$ as a 
function of $\mu_{x}$ and $\mu_{y}$ for $\mu_{s} = 0.1$ min$^{-1}$,  
$\mu_{s} = 0.5$ min$^{-1}$ and, $\mu_{s} = 1$ min$^{-1}$ in 
Fig.~\ref{fig7}(a-c), respectively, within multiplicative integration 
scheme. The bifurcation sub-motif dominates in all of these 
landscapes which are negatively valued in most of the parts. 
The domain with maximum negativity, shifts higher up the 
diagonal ($\mu_{x}=\mu_{y}$), as $\mu_{s}$ increases from 
0.1 min$^{-1}$ to 1 min$^{-1}$. The contribution from the 
integration sub-motif dominates mainly along the region where 
any one of the tuning parameters ($\mu_{x}$ or $\mu_{y}$) is 
small in comparison with its counterpart
while the remaining one scans its full range. As $\mu_{s}$ 
increases, the relative contribution made by the integration 
sub-motif increases.

In support of the difference in the net synergy shown in 
Fig.~\ref{fig7}(a-c), we now look at the difference in the SNR-s 
of the sub-motifs due to multiplicative integration scheme. 
The panels of Fig.~\ref{fig7}(d-f), depict the corresponding variation 
in the difference between SNR-s of integration and bifurcation 
sub-motifs, i.e. SNR$_{IM-DM}^{Mult}$-SNR$_{BM-DM}$ with 
$\mu_{s}$ taking values of $0.1$ min$^{-1}$, $0.5$ min$^{-1}$ 
and, $1$ min$^{-1}$, respectively. For computing SNR$_{BM-DM}$ 
and SNR$_{IM-DM}^{Mult}$, we have considered the 
signaling branches S $\rightarrow$ X and X $\rightarrow$ Z, 
respectively. We note that with $\mu_{s}=0.1$ min$^{-1}$, the 
map consists of both positive and negative valued domains. In 
the positive region, the IM-DM has higher fidelity than the BM-DM 
whereas, for some specific combinations of the tuning parameters, 
the BM-DM overpowers the IM-DM, in its efficiency in high fidelity 
information processing, thereby creating a negatively valued region 
in this map. As $\mu_{s}$ increases, the IM-DM takes control over 
the entire $(\mu_{x},\mu_{y})$ parameter space but the difference 
between the two SNR-s gradually decreases. This analysis 
graphically marks different regions where particular sub-motif plays the  
leading role compared to its counterpart in increasing the fidelity in 
information transmission.

\begin{figure}[!t]
\begin{center}
\includegraphics[width=1.0\columnwidth,angle=0]{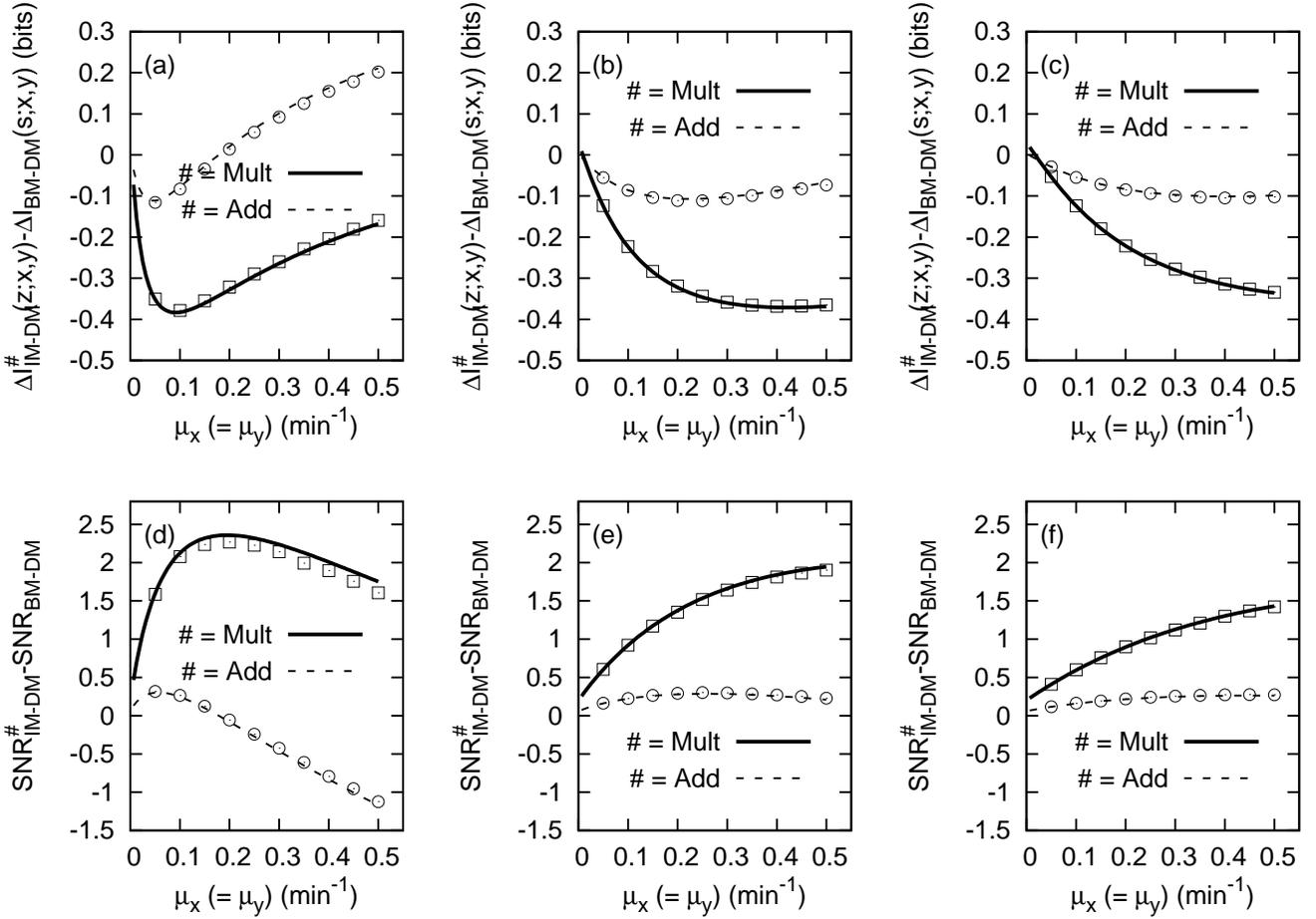}
\end{center}
\caption{Panels (a)-(c) depict $\Delta I_{IM-DM}^{Mult}(z;x,y)-\Delta I_{BM-DM}(s;x,y)$ and $\Delta I_{IM-DM}^{Add}(z;x,y)-\Delta I_{BM-DM}(s;x,y)$ (in bits) as functions of $\mu_{x}(=\mu_{y})$ for $\mu_{s}=$ 0.1, 0.5 and, 1 min$^{-1}$, respectively. Similarly, panels (d)-(f) show SNR$_{IM-DM}^{Mult}-$SNR$_{BM-DM}$ and SNR$_{IM-DM}^{Add}-$SNR$_{BM-DM}$ for $\mu_{s}=$ 0.1, 0.5 and, 1 min$^{-1}$, respectively. SNR$_{BM-DM}$ and SNR$_{IM-DM}^{Mult}$ (SNR$_{IM-DM}^{Add}$) are computed considering the signaling pathways S $\rightarrow$ X and X $\rightarrow$ Z, respectively. The lines are due to analytical results and the symbols represent numerical data generated using Gillespie's algorithm \cite{Gillespie1976,Gillespie1977} with an ensemble averaging over $10^6$ independent time series.The expressions are obtained using constant populations (at steady state) of the biochemical species: $\langle s \rangle=10$, $\langle x \rangle=100$, $\langle y \rangle=100$ and, $\langle z \rangle=100$ all in molecules/$V$ with $V$ being the unit effective cellular volume. The expressions for the rate parameters are $k_s$ = $\mu_s \langle s \rangle$, $k_x$ = $\mu_x \langle x \rangle ((K_1^n+\langle s \rangle^n)/\langle s \rangle^n)$, 
$k_y$ = $\mu_y \langle y \rangle ((K_2^n+\langle s \rangle^n)/\langle s \rangle^n)$,
$k_z$ = $\mu_z \langle z \rangle (\frac{\langle x \rangle^n}{K_3^n+\langle x \rangle^n}+\frac{\langle y \rangle^n}{K_4^n+\langle y \rangle^n})^{-1}$ and, 
$k_z'$ = $\mu_z \langle z \rangle 
((K_3^n+\langle x \rangle^n)/\langle x \rangle^n)
((K_4^n+\langle y \rangle^n)/\langle y \rangle^n)$. We have kept $K_1=K_2=90$, $K_3=K_4=100$ all in molecules/$V$ with $V$ being the unit effective cellular volume, $\mu_{z}=5$ min$^{-1}$ and, $n=1$ throughout to generate the profiles.}
\label{fig8}
\end{figure}

The two branches of both bifurcation and integration sub-motifs 
become identical when X and Y relax identically ($\mu_{x} = \mu_{y}$). 
Under such a condition, where the two pathways namely 
S$\rightarrow$X$\rightarrow$Z and, S$\rightarrow$Y$\rightarrow$Z 
are mirror images of each other, we explored in Fig.~\ref{fig8}(a-f)
how the differences in between the net synergy-s and SNR-s of sub-motifs 
respond to change in integration logic. The first observation is that the 
net synergy differences are mostly negatively constrained while 
SNR differences are mostly positively constrained. However, for low 
$\mu_{s}$ and for additive integration scheme these metrics span in 
both positive and negative spaces. For all strengths of the signal, it is 
observed that multiplicative integration gives more negative net synergy 
difference than additive integration. While for SNR differences, 
multiplicative integration contributes more positively than its additive 
counterpart. Two things, in particular, are worth mentioning. Mechanism 
wise, additive integration is seen to be spanning both positive and 
negative domains of the net synergy difference and SNR difference. 
The gap between contributions from multiplicative and additive 
mechanisms is the largest among all for the low $\mu_{s}$ case. 
This type of computational investigation scanning different regimes of 
the net synergy and SNR may be helpful while constructing synthetic 
biological circuits which often implement logical computations in 
decision making processes \cite{Silva-Rocha2008}. After looking at the strong and weak 
regimes and corresponding transitions, one may infer biologically 
plausible parameter sets with which synthetic circuits can be 
constructed to achieve specific goals, e.g. large net synergy and 
SNR differences. After that, the predictions can be tested against 
the output data generated through real experiments.   


\section{Conclusion}

To summarize, we have developed an information-theoretic characterization 
of DM using the formalism of PID. To know how much predictive 
power one biochemical component has about others, we have 
chosen the metric of net synergy that takes care of synergy, 
redundancy and, information independence. We have identified two 
sub-motifs namely the signal bifurcation motif 
and the signal integration motif out of the whole motif, i.e. the DM. From the set of
Langevin equations, using LNA and after that Lyapunov equation at steady state
we obtained the analytic expressions for the second moments of the Gaussian 
random variables. These variables represent different biochemical species of the 
network. The generic regulatory functions have been chosen to 
be nonlinear which is typically the case in real biological systems. In 
our analysis, we investigate the effect of the relaxation time scale on the 
information transmission in network motifs. It is important to note that 
a slower downstream species fails to sense the faster upstream 
species and thus, information propagation is hindered 
in the channel. This study attempts to depict the effect of variation in 
relaxation time scale in the net synergy landscape. Effective redundancy is lowered 
whenever upstream species fluctuates on a faster time scale compared 
to the downstream species. Apart from the time scales, signal integration 
mechanisms were observed to play a definitive role to nurture the net 
synergy. Our computations reveal that SNR in the DM increases with 
effective redundancy $I_{ER}$ for both signal integration logic.

Our study also uncovers the physical reason behind the creation of 
effective redundancy $I_{ER}$ in the net synergy profile. We observe 
only effective synergy $I_{ES}$ in the integration motif whose two 
signal integrating branches are architecturally disjoint from above, i.e. 
having no common source of fluctuations. Therefore, we are led to 
speculate that effective redundancy arises out of information sharing between 
different target nodes of the network, caused by a common source of 
fluctuations. Multiplicative integration of signal at the output is observed 
to generate more $I_{ER}$ than additive integration in IM-DM. 
In an IM, multiplicative integration logic produces more $I_{ES}$ than 
its additive counterpart.

There are indications emanating from variations of the net synergy 
as a function of various activation coefficients that biological motifs 
with better noise handling capabilities will prefer weak activation for 
intermediate information nodes and moderate activation for the final 
downstream target species for multiplicative integration 
scheme. For additive integration logic, the final downstream target also 
needs to be weakly activated by the intermediates which themselves 
are weakly activated by the source.

We quantified the relative information processing 
strengths of the BM-DM and, IM-DM operating under low, medium 
and, high relaxation frequency of the signal with both additive and 
multiplicative integration logic. The related maps portraying the 
interplay of SNR-s of the two sub-motifs, under similar parametric 
conditions in the $\mu_{x},\mu_{y}$ domain help us to identify 
regions where specific sub-motifs contribute more towards high 
fidelity information transmission. While constructing synthetic 
biological logic circuits, one can easily select plausible parameter 
sets generated in these observations to achieve the desirable magnitude of 
net synergy and SNR. By durability and 
performance of these artificially constructed biological motifs, one 
can positively infer why only specific types of signal processing 
motifs and not all, are abundant in nature. These are highly 
interesting queries of evolutionary importance, leading to better 
understanding of functionality in relation with the architecture of living 
organisms.

In a nutshell, we investigated the connection between the emergence of effective synergy 
and effective redundancy and, the architectural or topological features of the 
diamond motif and its sub-motifs. For this purpose, we quantified the net 
synergy in our information-theoretic study. Our 
generalized analysis employing a biologically relevant 
model and parameters presents key concepts and adds novelty to 
the current understanding of information processing in a diamond 
motif. In future work, this information-theoretic framework as 
presented in this communication has enormous scope to reveal 
novel findings in other abundant network motifs. Eventually, it may 
lead towards the discovery of some unifying physical principles governing 
diversified biological systems constituted across length scales and 
operating across time scales.

 
\begin{acknowledgments}
Ayan Biswas is thankful to Bose Institute, Kolkata for a research fellowship.
Financial support from the Council of Scientific and Industrial Research (CSIR), India
[01(2771)/14/EMR-II] is thankfully acknowledged.
\end{acknowledgments}


\appendix 

\section{The bifurcation motif}

For a bifurcation motif, one needs to consider the dynamics of only S, X and, Y as expressed with the following set of Langevin equations
\begin{eqnarray*}
\frac{ds}{dt} & = & f_{s}(s)-\mu_s s+\xi_s(t), \\
\frac{dx}{dt} & = & f_{x}(s,x)-\mu_x x+\xi_x(t), \\
\frac{dy}{dt} & = & f_{y}(s,y)-\mu_y y+\xi_y(t).
\end{eqnarray*}

The corresponding Jacobian obtained from linearizing this set is as follows:
\begin{eqnarray*}
\mathbf{J}= \left( 
\begin{array}{ccc} f^{'}_{s,s} - \mu_{s} & 0 & 0 \\
f^{'}_{x,s} & f^{'}_{x,x} - \mu_{x} & 0 \\
f^{'}_{y,s} & 0 & f^{'}_{y,y} - \mu_{y}  \\
\end{array} \right),
\end{eqnarray*}

\noindent
where $f^{'}_{s,s} \equiv f^{'}_{s,s}(\langle s \rangle)$, $f^{'}_{x,s} \equiv f^{'}_{x,s}(\langle s \rangle, \langle x \rangle)$, $f^{'}_{x,x} \equiv f^{'}_{x,x}(\langle s \rangle, \langle x \rangle)$, $f^{'}_{y,s} \equiv f^{'}_{y,s}(\langle s \rangle, \langle y \rangle)$ and, $f^{'}_{y,y} \equiv f^{'}_{y,y}(\langle s \rangle, \langle y \rangle)$.
Here, $\langle \cdots \rangle$ denotes the steady state ensemble average and $f^{'}_{s,s}(\langle s \rangle)$ symbolically means that the regulatory function $f_{s}$ has been differentiated with respect to $s$ and evaluated at $\langle s \rangle$ and so on. 
Using the Jacobian, we solve the Lyapunov equation (\ref{eq6}) and derive the analytic expressions for variance and covariance associated with the bifurcation motif. 
\begin{eqnarray}
\Sigma(s) & = & \frac{\alpha_{s}}{2(\mu_{s}-f^{'}_{s,s})}, \\
\Sigma(s,x) & = & \frac{f^{'}_{x,s} ~ \Sigma(s)}{(\mu_{s}-f^{'}_{s,s})+(\mu_{x}-f^{'}_{x,x})}, \\
\Sigma(s,y) & = & \frac{f^{'}_{y,s} ~ \Sigma(s)}{(\mu_{s}-f^{'}_{s,s})+(\mu_{y}-f^{'}_{y,y})}, \\
\Sigma(x) & = & \frac{\alpha_{x}}{2(\mu_{x}-f^{'}_{x,x})} + \frac{f^{'}_{x,s} ~ \Sigma(s,x)}{(\mu_{x}-f^{'}_{x,x})}, \\
\Sigma(y) & = & \frac{\alpha_{y}}{2(\mu_{y}-f^{'}_{y,y})}+\frac{f^{'}_{y,s} ~ \Sigma(s,y)}{(\mu_{y}-f^{'}_{y,y})}, \\
\Sigma(x,y) & = & \frac{f^{'}_{y,s} ~ \Sigma(s,x)+f^{'}_{x,s} ~ \Sigma(s,y)}{(\mu_{x}-f^{'}_{x,x})+(\mu_{y} - f^{'}_{y,y})}. 
\end{eqnarray}

\noindent
In our calculation, we have used $f_{s}$ = $k_{s}$, $f_{x}$ = $k_{x} (s^n/(K_{1}^n+s^n))$ and,  $f_{y}$ = $k_{y} (s^n/(K_{2}^n+s^n))$.  $\alpha_{i} \equiv \langle |\xi_{i}|^{2}\rangle$ $(i = s,x,y)$ imply the ensemble averaged noise strengths evaluated at steady state.

\section{The integration motif}

For the integration motif, only species X, Y and, Z are taken into account for which the Langevin description stands like the following:
\begin{eqnarray*}
\frac{dx}{dt} & = & f_{x}(x)-\mu_x x+\xi_x(t), \\
\frac{dy}{dt} & = & f_{y}(y)-\mu_y y+\xi_y(t), \\
\frac{dz}{dt} & = & f_{z}(x,y,z)-\mu_z z+\xi_z(t).
\end{eqnarray*}

\noindent with the Jacobian,
\begin{eqnarray*}
\mathbf{J}= \left( 
\begin{array}{ccc} f^{'}_{x,x} - \mu_{x} & 0 & 0 \\
0 & f^{'}_{y,y} - \mu_{y} & 0 \\
f^{'}_{z,x} & f^{'}_{z,y} & f^{'}_{z,z} - \mu_{z}  
\end{array} \right).
\end{eqnarray*}

\noindent 
Here $f^{'}_{x,x} \equiv f^{'}_{x,x}( \langle x \rangle)$, $f^{'}_{y,y} \equiv f^{'}_{y,y}( \langle y \rangle)$,
$f^{'}_{z,x} \equiv f^{'}_{z,x}( \langle x \rangle, \langle y \rangle, \langle z \rangle)$, $f^{'}_{z,y} \equiv f^{'}_{z,y}(\langle x \rangle, \langle y \rangle, \langle z \rangle)$ and, $f^{'}_{z,z} \equiv f^{'}_{z,z}( \langle x \rangle, \langle y \rangle, \langle z \rangle)$.
As in the previous case $\langle \cdots \rangle$ denotes a steady state ensemble average and $f^{'}_{x,x}(\langle x \rangle)$ symbolically means that the regulatory function $f_{x}$ has been differentiated with respect to $x$ and evaluated at $\langle x \rangle$, and so on.
Solving the corresponding Lyapunov equation (\ref{eq6}) at steady state yields expressions of the second moments,
\begin{eqnarray}
\Sigma(x) & = & \frac{\alpha_{x}}{2(\mu_{x}-f^{'}_{x,x})}, \\
\Sigma(y) & = & \frac{\alpha_{y}}{2(\mu_{y}-f^{'}_{y,y})}, \\
\Sigma(x,y) & = & 0, \\
\Sigma(x,z) & = & \frac{f^{'}_{z,x} ~ \Sigma(x)}{(\mu_{x}-f^{'}_{x,x}) + (\mu_{z} - f^{'}_{z,z})}, \\
\Sigma(y,z) & = & \frac{f^{'}_{z,y} ~ \Sigma(y)}{(\mu_{y}-f^{'}_{y,y})+(\mu_{z}-f^{'}_{z,z})}, \\
\Sigma(z) & = & \frac{\alpha_{z}}{2(\mu_{z}-f^{'}_{z,z})} \nonumber \\
&& +\frac{f^{'}_{z,x} ~ \Sigma(x,z) + f^{'}_{z,y} ~ \Sigma(y,z)}{(\mu_{z}-f^{'}_{z,z})}.
\end{eqnarray}

\noindent
Here, $f_{x}$ = $k_{x}$, $f_{y}$ = $k_{y}$, $f_{z}$ = $k_{z} ((x^n/(K_{3}^n+x^n))+(y^n/(K_{4}^n+y^n)))$ (Additive integration), $f_{z}$ = $k_{z}' (x^n/(K_{3}^n+x^n)) (y^n/(K_{4}^n+y^n))$ (Multiplicative integration) and, $\alpha_{i}$-s $(i = x,y,z)$ imply, as in the previous case, steady state ensemble averaged noise strengths of different biochemical species.

\section{The diamond motif}

For a diamond motif where S, X, Y and, Z are all involved, the full set of Langevin equations are
\begin{eqnarray*}
\frac{ds}{dt} & = & f_{s}(s)-\mu_s s+\xi_s(t), \\
\frac{dx}{dt} & = & f_{x}(s,x)-\mu_x x+\xi_x(t), \\
\frac{dy}{dt} & = & f_{y}(s,y)-\mu_y y+\xi_y(t), \\
\frac{dz}{dt} & = & f_{z}(s,x,y,z)-\mu_z z+\xi_z(t).
\end{eqnarray*}

\noindent For the above-mentioned kinetics, the Jacobian becomes
\begin{eqnarray*}
\mathbf{J}= \left( 
\begin{array}{cccc} f^{'}_{s,s} - \mu_{s} & 0 & 0 & 0 \\
f^{'}_{x,s} & f^{'}_{x,x} - \mu_{x} & 0 & 0 \\
f^{'}_{y,s} & 0 & f^{'}_{y,y} - \mu_{y} & 0 \\
f^{'}_{z,s} & f^{'}_{z,x} & f^{'}_{z,y} & f^{'}_{z,z} - \mu_{z}  
\end{array} \right). \nonumber \\
\end{eqnarray*}

\noindent
Here, $f^{'}_{s,s} \equiv f^{'}_{s,s}(\langle s \rangle)$, $f^{'}_{x,s} \equiv f^{'}_{x,s}(\langle s \rangle, \langle x \rangle)$, $f^{'}_{x,x} \equiv f^{'}_{x,x}(\langle s \rangle, \langle x \rangle)$, $f^{'}_{y,s} \equiv f^{'}_{y,s}(\langle s \rangle, \langle y \rangle)$, $f^{'}_{y,y} \equiv f^{'}_{y,y}(\langle s \rangle, \langle y \rangle)$, $f^{'}_{z,s} \equiv f^{'}_{z,s}(\langle s \rangle, \langle x \rangle, \langle y \rangle, \langle z \rangle)$, $f^{'}_{z,x} \equiv f^{'}_{z,x}(\langle s \rangle, \langle x \rangle, \langle y \rangle, \langle z \rangle)$, $f^{'}_{z,y}\equiv f^{'}_{z,y}(\langle s \rangle, \langle x \rangle, \langle y \rangle, \langle z \rangle)$ and, $f^{'}_{z,z} \equiv f^{'}_{z,z}(\langle s \rangle, \langle x \rangle, \langle y \rangle, \langle z \rangle)$.
The notations $\langle \cdots \rangle$, $f^{'}_{s,s}(\langle s \rangle)$, etc. have the usual meaning as mentioned in the previous two cases.
 Again with the help of the Lyapunov equation (\ref{eq6}), we get the following analytic expressions for variance and covariance,
\begin{eqnarray}
\Sigma(s) & = & \frac{\alpha_{s}}{2(\mu_{s}-f^{'}_{s,s})}, \\
\Sigma(s,x) & = & \frac{f^{'}_{x,s} ~ \Sigma(s)}{(\mu_{s}-f^{'}_{s,s})+(\mu_{x}-f^{'}_{x,x})}, \\
\Sigma(s,y) & = & \frac{f^{'}_{y,s} ~ \Sigma(s)}{(\mu_{s}-f^{'}_{s,s})+(\mu_{y}-f^{'}_{y,y})}, \\
\Sigma(s,z) & = & \frac{f^{'}_{z,s} ~ \Sigma(s) +
f^{'}_{z,x} ~ \Sigma(s,x) +
f^{'}_{z,y} ~ \Sigma(s,y)
}{
(\mu_{s}-f^{'}_{s,s})+(\mu_{z}-f^{'}_{z,z}) 
}, \\
\Sigma(x) & = & \frac{\alpha_{x}}{2(\mu_{x}-f^{'}_{x,x})}+\frac{f^{'}_{x,s} ~ \Sigma(s,x)}{(\mu_{x}-f^{'}_{x,x})}, \\
\Sigma(y) & = & \frac{\alpha_{y}}{2(\mu_{y}-f^{'}_{y,y})}+\frac{f^{'}_{y,s} ~ \Sigma(s,y)}{(\mu_{y}-f^{'}_{y,y})}, \\
\Sigma(x,y) & = & \frac{f^{'}_{x,s} ~ \Sigma(s,y)+f^{'}_{y,s} ~ \Sigma(s,x)}{(\mu_{x}-f^{'}_{x,x})+(\mu_{y}-f^{'}_{y,y})}, \\
\Sigma(x,z) & = & \frac{f^{'}_{x,s} \Sigma(s,z) + 
f^{'}_{z,s} \Sigma(s,x) +
f^{'}_{z,x} \Sigma(x) + 
f^{'}_{z,y} \Sigma(x,y)
}{
(\mu_{x}-f^{'}_{x,x}) + (\mu_{z}-f^{'}_{z,z})
}, \nonumber \\
&& \\
\Sigma(y,z) & = & \frac{
f^{'}_{y,s} \Sigma(s,z) +
f^{'}_{z,s} \Sigma(s,y) +
f^{'}_{z,y} \Sigma(y) +
f^{'}_{z,x} \Sigma(x,y)
}{
(\mu_{y}-f^{'}_{y,y}) +(\mu_{z}-f^{'}_{z,z})
}, \nonumber \\
&& \\
\Sigma(z) &= & \frac{\alpha_{z}}{2(\mu_{z}-f^{'}_{z,z})} \nonumber \\
&& +\frac{f^{'}_{z,s} \Sigma(s,z)+f^{'}_{z,x} \Sigma(x,z)+f^{'}_{z,y} \Sigma(y,z)}{(\mu_{z}-f^{'}_{z,z})}.
 \end{eqnarray}

For diamond motif, the regulatory functions are chosen as, $f_{s}$ = $k_{s}$,  $f_{x}$ = $k_{x} (s^n/(K_{1}^n+s^n))$,  $f_{y}$ = $k_{y} (s^n/(K_{2}^n+s^n))$, $f_{z}$ = $k_{z} ((x^n/(K_{3}^n+x^n))+(y^n/(K_{4}^n+y^n)))$ (Additive integration), $f_{z}$ = $k_{z}' (x^n/(K_{3}^n+x^n)) (y^n/(K_{4}^n+y^n))$ (Multiplicative integration) and $\alpha_{i}$-s $(i = s,x,y,z)$ stand for different steady state noise strengths obtained through ensemble averaging.  
We utilize these expressions of second moments (Eqs.~C1-C10) to compute $\Delta I_{BM-DM} (s;x,y)$ and $\Delta I_{IM-DM} (z;x,y)$ (for additive and multiplicative integration scheme), as required.


\end{document}